\RequirePackage{tkz-euclide}
\RequirePackage{pgfplots}
\pgfplotsset{compat=1.11}
\usepgfplotslibrary{fillbetween}
\RequirePackage{tikz}
\usetikzlibrary{shapes,arrows,calc,positioning,automata,intersections,plotmarks,quotes,angles}
\usetikzlibrary{positioning}
\usetikzlibrary{decorations.pathmorphing} 
\usetikzlibrary{matrix,fit} 
\tikzstyle{block} = [draw, rectangle, thick, text centered, minimum height=0.7cm, minimum width=1cm]
\tikzstyle{sum} = [draw, thick, circle, node distance=1.5cm] 

\documentclass[onecolumn,a4paper]{article}

\usepackage[english]{babel}
\usepackage{amsthm,amsmath,amssymb,amsfonts,amsopn}
\usepackage{color}
\usepackage{graphicx,subfigure,graphics,epstopdf,float}
\graphicspath{{figures/}}
\usepackage{enumerate}
\usepackage{algorithmic}
\usepackage{makecell}
\usepackage{booktabs}
\usepackage{hyperref}
\usepackage{soul}
\usepackage{stmaryrd}

\newcommand{\E}{\mathrm{\bf E}}

\newcommand{\Vt}{\tilde{V}}
\newcommand{\Ut}{\tilde{U}}
\newcommand{\Mt}{\tilde{M}}
\newcommand{\Nt}{\tilde{N}}

\newcommand{\norm}[2]{\|#1\|_{#2}}
\newcommand{\mnorm}[2]{\interleave #1\interleave_{#2}}

\newcommand{\rh}[1]{\mathbb{RH}_{#1}}

\newtheorem{theorem}{Theorem}
\newtheorem{corollary}{Corollary}%
\newtheorem{lma}{Lemma}%
\newtheorem{assump}{Assumption}%
\newtheorem{remark}{Remark}%
\newtheorem{definition}{Definition}%

\begin{document}

\title{Mean-square stability of linear systems over channels with random transmission delays}
\author{Jieying Lu\thanks{School of Automation Science and Engineering, South China University of Technology,
Guangzhou, China. \href{mailto:aujylu@scut.edu.cn}{aujylu@scut.edu.cn},
\href{mailto:l.junhui@mail.scut.edu.cn}{l.junhui@mail.scut.edu.cn},
\href{mailto:wzhsu@scut.edu.cn}{wzhsu@scut.edu.cn}}
\and Junhui Li\footnotemark[1]
\and Weizhou Su \footnotemark[1]}
\date{}

\maketitle

\begin{abstract}
This work studies the mean-square stability and stabilization problem for networked feedback systems. Data transmission delays in the network channels of the systems are considered.
It is assumed that these delays are  i.i.d. processes with given probability mass functions (PMFs).
A necessary and sufficient condition of mean-square (input-output) stability is studied for the networked feedback systems in terms of the input-output model and state-space model. Furthermore, according to this condition, mean-square stabilization via output feedback is studied for the networked feedback systems.
\end{abstract}


\section{Introduction}\label{Sec:Introduction}

Recently, it is known that data transmission delay is a major network-induced phenomenon that affects the control performance and may even lead to instability of networked control systems (NCSs) \cite{Antsaklis2007}.
For about two decades, there have been results reported for the stability and stabilization of discrete-time NCSs in the presence of data transmission delays (referred to as network-induced delays usually, see, e.g., \cite{Gomez2013Stability,Liang2017Optimal,QuevedoJ_TAC2014, Zhang2015Stabilization}).
As widely recognized, network-induced delays can be time-varying and possibly random, unlike these studied in traditional researches wherein systems are with constant input delays, see experiments in \cite{Tsang2003Network}.
For random network-induced delays of discrete-time NCSs, there are mainly two models, namely the Markov-based delays and the independent and identically distributed (i.i.d.) transmission delays.
NCSs with the former delays are usually converted into Markovian jump systems such that the stabilization conditions can be obtained in terms of a set of linear matrix inequalities (LMIs) with nonconvex constraints (see  for example \cite{Zhang2005A}). Hence,
for NCSs with time-varying network-induced delays, most results in stabilization design derived based on the construction of Lyapunov functionals were only sufficient conditions.
(see, e.g., \cite{Cloosterman2009Stability,Zhang2016NewStability}).
In a series of more recent works (\cite{ Dipankar2022Optimal, QuevedoJ_TAC2014, Schenato2008Optimal, SLL2021, Su2017mean-square}), it was shown that the i.i.d. transmission delay model provides suitable and practical framework to model the multi-path transmission phenomenon.
For NCSs with i.i.d. transmission delays, most research works focus on state estimation and (finite-horizon) optimal control problems (see, e.g., \cite{Dipankar2022Optimal, Schenato2008Optimal}), while, in \cite{QuevedoJ_TAC2014}, the stabilization problem via sequence-based control was studied for a nonlinear networked system with i.i.d. transmission delays and only a sufficient condition was obtained.
The small gain theorem of the mean-square input-output stability with these delays is studied for the networked systems in \cite{SLL2021, Su2017mean-square}.

This paper aims to study the mean-square stability and stabilization problem for NCSs with i.i.d. transmission delays, and to establish the connection between mean-square input-output stability and mean-square stability of the networked systems.
To this end, an analytic description is discussed for characterizing data transmission processes with  random transmission  delays. With this description, an input-output model of the channel is presented in terms of its impulse response. It turns out that this analytic channel model can be decomposed into a deterministic mean channel and a channel uncertainty induced by the random delays. By virtue of the decomposition, a small gain-type necessary and sufficient condition for the mean-square input-output stability of the NCS with random transmission delays is obtained in the frequency domain, inspired by the seminal work of \cite{Willems1971Frequency}. Subsequently, it is found that the output of the system is asymptotically stationary when its input is asymptotically stationary and the system is mean-square input-output stable. The connection between mean-square input-output stability and mean-square stability of the networked systems is established. Finally, an optimal design
for the mean-square stabilization via output feedback in state-space model is studied.

The rest of this paper is organized as follows.
Section \ref{Sec:Problem_Formulation} characterizes the random transmission delays of the unreliable channel and formulates the mean-square input-output stability and stabilization problem. Section \ref{Sec:Model_A} discusses the input-output model and second-order statistics of the channel uncertainty induced by random transmission delays.
Section \ref{Sec:MSS} presents the main results of mean-square stability and stabilization via output feedback for the closed-loop system. 
Section \ref{Sec:MSS_output_feedback} studies the mean-square stabization design via output feedback. 
A numerical example is studied in Section \ref{section_example}.
Section \ref{Sec:Conclusion} draws conclusions. 

The notations used in this paper is mostly standard.
$A^*, A^{-1}, A^{-*}, A^T$ are, respectively, the complex conjugate transpose, inverse, inverse conjugate transpose, and transpose of appropriate matrix $A$.
$\rh{\infty}$ is the set of all proper stable rational transfer functions.
Denote the $\mathbb{H}_2$ norm of a discrete-time proper rational transfer function $G(z)$ by ${\left\| G(z) \right\|_2}$ (see \cite{Chen1995Optimal} for details),
$$
\|G(z)\|_2^2=\frac{1}{2\pi}\int_{-\infty}^{\infty} G^*(e^{j\theta})G(e^{j\theta}) d\theta.
$$
Furthermore, $\mathbb{R}$ stands for the set of real numbers.
$\E\{\cdot\}$ denotes the expectation operator of a random variable, and $\delta(\cdot)$ denotes the Kronecker delta function that $\delta \left( 0 \right) = 1$ and $\delta \left( i \right) = 0$ for any integer $i \ne 0$.

\section{Problem Formulation}
\label{Sec:Problem_Formulation}

\begin{figure}[hbt]
  \centering
\begin{tikzpicture}[auto, node distance=2cm, >=stealth', line width=0.75pt]
\node[sum] (sum) {};
\node[block, right of= sum, xshift=-0.3cm] (G) {$P$};
\node[block, below of= G, xshift=0cm] (K) {$K$};
\node[block, below of= sum, yshift=0.9cm, minimum width= 1.2cm, minimum height=0.6cm] (omega) {$\Delta$};
\draw[->] ($(sum.west) + (-1,0)$) -- node[near start](){$v$} (sum);
\draw[->] (sum) -- (G);
\draw[->] (G) -- node[near end]{$y$} ($(G)+(2.4cm,0cm)$);
\draw[->] (G) -- ++(1.6cm,0cm) |- (K);
\draw[->] (K) -| node[]{$u$} (omega);
\draw[->] (omega) -- node[near start]{$u_d$} node[pos=0.8]{$-$} (sum);
\end{tikzpicture}
\caption{An LTI system over a channel with random transmission delays}
\label{Fig:System_with_uncertainty}
\end{figure}
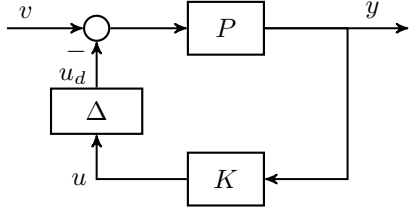

In this work, a networked feedback system over an unreliable channel, as shown in Fig. \ref{Fig:System_with_uncertainty}, is considered where $P$ is a single-input and single-output plant which is a linear time-invariant (LTI) system with relative degree greater than zero,  $K$ is a controller, and $\Delta$ is an unreliable channel.
The signals $v,y$ and $u$ are the external input, the plant output, and the control signal, respectively, while the signal $u_d$ stands for the corrupted control signal received by the terminal of the unreliable channel. The signals  $v,y, u$ and $u_d$ are from $\mathbb{R}$.

\subsection{The unreliable channel with random transmission delays}

Denote the time which is used in transmitting data $u(n)$ sent at the instant $n$ over the unreliable channel
$\Delta$ by $\tau_n$. That is, the data $u(n)$ transmitted over the channel $\Delta$ at the instant $n$ is received at the instant $n+\tau_n$ by the terminal of the channel.
\begin{assump}\label{ass1}
The sequence $\left\{  \tau_n \right\}$ is an i.i.d process.  For a given $n$,  the random variable $\tau_n$ is from the set ${\cal D}=\left\{0,1,2, \cdots, \tau \right\}$ where $ \tau$ is the largest possible delay step. The random variable $\tau_n$ is of a given probability mass function (PMF), i.e.,
\begin{align*}
Pr\left\{ \tau_n=i  \right\}=p_{i}, i=0,1,\cdots, \tau
\end{align*}
and
\begin{align*}
\sum_{i=0}^{ \tau}p_{i}=1.
\end{align*}
\end{assump}
In Assumption \ref{ass1}, it is assumed that the largest delay step in the channel $\Delta$ is $\tau$,  all possible data received by the terminal of $\Delta$ at the instant $k$ include $u(k), u(k-1), \cdots, u(k-\tau)$. The times used in transmitting these data are $\tau_k$, $\tau_{k-1}$, $\cdots$,  $\tau_{k-\tau}$, respectively, which are random variables. The functions $\delta(\tau_k)$, $\delta(\tau_{k-1}-1)$, $\cdots$,
$\delta(\tau_{k-\tau}-\tau)$ indicate whether the data $u(k), u(k-1), \cdots, u(k-\tau)$ are received by the terminal of $\Delta$ at instant $k$, respectively. For the received data, it is assumed that the transmitted data have time-stamps. A linear combination of the data received by the terminal of $\Delta$ at the instant $k$ is used as its output, i.e.,
\begin{align}\label{input-output_j}
u_{d}(k)=\sum_{i=0}^{\tau}\alpha_{i}\delta(\tau_{k-i}-i)u(k-i).
\end{align}
where $\alpha_{0}$, $\alpha_{1}, \cdots$, $\alpha_{\tau}$  are a set of given weights assigned to the received data in terms of the data's delay steps.

The equation (\ref{input-output_j}) describes the input-output relation of $\Delta$. Thus, it is clear that $\Delta$ is a causal linear time-varying system with a finite impulse response. According to this input-out relation, 
the unit-impulse response of $\Delta$ is given by
\begin{align}\label{Equ:FIR_Delta}
h(k,n) &= \left\{ \begin{array}{ll}
						0, & k<n \\
                        \alpha_{k-n}\delta(\tau_n-(k-n)) ,& n \leq  k \leq n+\tau\\
                                      0, & k>n+\tau
                        \end{array} \right.
\end{align}
where $n$ is the instant when the impulse is applied. Furthermore, the response of $\Delta$ to its unit
impulse $\delta(k-n)$ input signal
is a finite length sequence $\left\{\alpha_{0}\delta(\tau_n), \alpha_{1}\delta(\tau_n-1)\right.$,
 $\left. \cdots, \alpha_{\tau}\delta(\tau_n-\tau) \right\}$ if the channel is at rest when $k=n$.

\subsection{Channel uncertainty induced by of random transmission delays}

It follows from Assumption \ref{ass1} that:
\begin{align*}
\E \{ \delta(\tau_n-i) \} =p_{i},\;\; i=0,1, \cdots, \tau.
\end{align*}
The mean sequence of the response $\left\{\alpha_{0}\delta(\tau_n)\right.$,
$\alpha_{1}\delta(\tau_n-1)$, $\cdots$, $\left. \alpha_{\tau}\delta(\tau_n-\tau)   \right\}$
of $\Delta$ to the unit impulse input $\delta(k-n)$ is given by
$\left\{\alpha_{0} p_{0}, \alpha_{1} p_{1}\right.$,
 $\left. \cdots, \alpha_{\tau} p_{\tau} \right\}$. Since this mean sequence describes the input-output relation of the channel in average sense, the channel whose unit impulse response is the mean sequence is referred to as  the mean channel and denoted by $H$. Under Assumption \ref{ass1}, the unit impulse response of the mean channel is
of the time-shift property (see page 32, \cite{ChenT1999}), so it is an LTI system. 
The transfer function of $H$ is given by
\begin{align}\label{H_j}
H(z)=\alpha_{0} p_{0}+ \alpha_{1} p_{1} z^{-1} + \cdots + \alpha_{\tau} p_{ \tau} z^{-\tau}.
\end{align}
Denote the output of $H$ by $\bar{u}(k)$, i.e.,
\begin{align}\label{H_j_output}
\bar{u}(k) &= \sum_{i= 0}^{\tau} \alpha_{i} p_{i} u(k-i). 
\end{align}

On the other hand, the deviation of $h(k,n)$ is given by
\begin{align*}
\omega(k,n)=h(k,n)-\E \{h(k, n)\}.
\end{align*}
From (\ref{Equ:FIR_Delta}) and Assumption \ref{ass1}, it holds that
\begin{align}\label{Equ:omega_Delta}
\omega(k,n) &= \left\{ \begin{array}{ll}
						0, & k<n \\
                        \alpha_{k-n}[\delta(\tau_n-(k-n))-p_{k-n}] ,& n \leq k \leq n+\tau\\
                                     0, &  k>n+\tau.
                        \end{array} \right.
\end{align}
It is clear that for any given $n$, the sequence $\left\{\omega(k,n)\right\}$ is dependent to $\tau_n$ only and describes 
the deviation $\Delta-H$ of the channel $\Delta$ induced by random transmission delay $\tau_n$.
In general, this deviation is referred to as the channel uncertainty and denoted by
$\Omega$. Denote the output of $\Omega$ by $d$, then it follows from the definition of $\omega(k,n)$ that
\begin{align}\label{omega_j_output}
d(k)=\sum_{i=0}^{\tau}\omega(k,k-i)u(k-i).
\end{align}

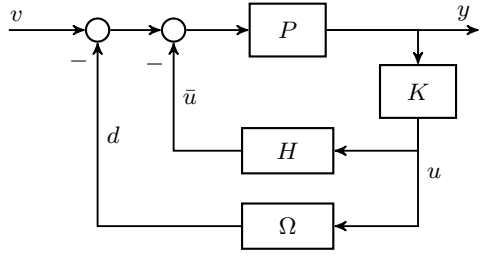
\begin{figure}[!hbt]
  \centering
\begin{tikzpicture}[auto, node distance=2cm, >=stealth', line width=0.7pt]
\small
\node[block](P){$P$};
\node[sum, left of= P, xshift=0cm](sum1){};
\node[sum, left of= sum1, xshift=0.5cm](sum2){};
\node[block, below right of= P, xshift=0.3cm, yshift=0.6cm](K){$K$};
\node[block, below of= P, yshift=0.4cm, minimum width= 1.2cm, minimum height=0.6cm](H){$H$};
\node[block, below of= H, yshift=1cm, minimum width= 1.2cm, minimum height=0.6cm](D){$\Omega$};

\draw[->]($(sum2.west)+(-1cm,0cm)$) -- node[pos=0.1]{$v$} (sum2);
\draw[->](sum2) -- (sum1);
\draw[->](sum1) -- node[]{} (P);
\draw[->](P) -| (K);
\draw[->](P) --  node[pos=0.9]{$y$} ($(P.east)+(2cm,0cm)$);
\draw[->](K) |- (H);
\draw[->](H) -| node[near end, swap]{$\bar{u}$} node[pos=0.9]{$-$} (sum1);
\draw[->](K) |-  node[near start]{$u$} (D);
\draw[->](D) -| node[near end, swap]{$d$} node[pos=0.95]{$-$} (sum2);
\end{tikzpicture}
  \caption{Equivalent framework of the LTI system with random channel delays }\label{Fig:System_G_with_uncertainty}
\end{figure}

Taking account to (\ref{input-output_j}), (\ref{H_j_output}) and (\ref{omega_j_output}), one can see that the output of $\Delta$ is the sum of the outputs of its mean channel $H$ and channel uncertainty $\Omega$, i.e.,
\begin{align} \label{u_d_decomp}
u_{d}(k)=\bar{u}(k)+d(k).
\end{align}
As a result, the closed-loop system in Fig. \ref{Fig:System_with_uncertainty} can be re-diagrammed as that in Fig. \ref{Fig:System_G_with_uncertainty}.

From \eqref{Equ:omega_Delta},  the first- and second-order statistics of $\omega(k,n)$ are given by
the following lemma.
\begin{lma}\label{Lma:omega_i}
Suppose that the random transmission delay process $\{\tau_n: -\infty < n < \infty\}$ satisfies Assumption
\ref{ass1}. Then it holds to the impulse response $\omega(k,n)$ of the channel uncertainty
$\Omega$ that
\begin{enumerate}
\item
for $i \in \mathcal{D}$,
\begin{equation*}
\E\{\omega(k,k-i)\} = 0;
\end{equation*}
\item for $i \in \mathcal{D}$,
\begin{equation*}
\hspace{-0.3cm}
\E\left\{ \omega(k_1, k_1-i) \omega(k_2, k_2-i) \right\}= \delta (k_1 - k_2) \alpha_{i}^2 p_{i}{(1 - p_{i})};
\end{equation*}
\item for $i_1 \neq i_2$, $i_1,i_2 \in \mathcal{D}$,
\begin{align*}
&\E\left\{ \omega(k_1,k_1-i_1) \omega(k_2, k_2-i_2) \right\}
= - \delta (k_1-i_1-k_2+i_2) \alpha_{i_1} \alpha_{i_2} p_{i_1} p_{i_2}.
\end{align*}
\end{enumerate}
\end{lma}
Please see  \cite{SLL2021} for proof.
\begin{remark}
Lemma \ref{Lma:omega_i} shows that, for any given $n_1$ and $n_2$, if $n_1\neq n_2$, the sequences $\{\omega(k, n_1)\}$ and $\{\omega(k, n_2)\}$ are mutually independent of.
Lemma  \ref{Lma:omega_i}.2 presents the variances of all entries of the sequence
$\left\{ \omega(n, n), \omega(n+1, n), \cdots \right.$, $\left.\omega(n+\tau, n) \right\}$.
Lemma  \ref{Lma:omega_i}.3 presents the correlations of $ \omega(n+i_1, n)$ and $ \omega(n+i_2, n)$ for $i_1 \neq i_2$ $\in {\cal D}$ in the sequence.
Moreover, it is shown that the second-order  statistics of $\omega(k,n)$ are only
determined by the $k-n$. This implies that  the second-order statistics of this channel uncertainty are time-invariant.
\end{remark}

\subsection{Problem formulation}

Now, the problems under studied in this work are discussed.
The following assumption is mostly standard in the existing work, see, e.g., \cite{Qi2017Control}.
\begin{assump}\label{Assum:v}
For the system in Fig. \ref{Fig:System_G_with_uncertainty}, the data transmission process $\{\tau_n\}$ is independent of the random input sequence $\{v(k)\}$, and $v(k)$ is a white noise with zero-mean and bounded variance
$\sigma_v^2(k)$.
\end{assump}

Define the nominal system of the system in  Fig. \ref{Fig:System_G_with_uncertainty} by the system without 
the channel uncertainty $\Omega$. Denote the transfer function of the nominal system from $d$ to $u$ by $G(z)$,  i.e.,
\begin{align}
G(z) &= P(z)K(z)[I + P(z) K(z) H(z)]^{-1}. \label{Equ:Nominal_system_G(z)} 
\end{align}
For convenience, we re-diagram Fig. \ref{Fig:System_G_with_uncertainty} as Fig. \ref{Fig:G_Omega}, which is an interconnection of the nominal system $G(z)$ and the zero-mean uncertainty $\Omega$.
Let the set of all proper controllers stabilizing $G(z)$ be $\mathcal{K}$.
Throughout this paper, we focus on the mean-square input-output stability (see for example \cite{Qi2017Control}), mean-square stability (see for example \cite{Elia2005Remote}) and stabilizability via output feedback defined next.
\begin{definition}\label{Def:Internally_MS}
The system in Fig. \ref{Fig:G_Omega} is said to be mean-square input-output stable if the nominal system $G(z)$ is stable (i.e., its poles are within the open unit circle) and for any input sequence $\{v(k)\}$ satisfying Assumption \ref{Assum:v}, the variances of $d(k)$ and $u(k)$ are also bounded, i.e., $\E\{d^2(k)\}<\infty$ and $\E\{u^2(k)\}<\infty$.
\end{definition}
\begin{definition}\label{Def:MS}
The system in Fig. \ref{Fig:G_Omega} is said to be mean-square stable if the nominal system $G(z)$ is stable (i.e., its poles are within the open unit circle), its input sequence $\{v(k)\}$ is void and the initial state $x(0)$ of the nominal system $G(z)$ is a random vector which is independent of the transmission delay sequence $\left\{ \tau_n \right\}$, the co-variance of the state $x(k)$ of the nominal system is convergent to zero.
\end{definition}
\begin{definition}\label{Def:Mean_square_stabilizable}
The system in Fig. \ref{Fig:G_Omega} is said to be mean-square stabilizable via output feedback if there exists a feedback controller $K \in \mathcal{K}$ such that the closed-loop system is mean-square input-output stable.
\end{definition}
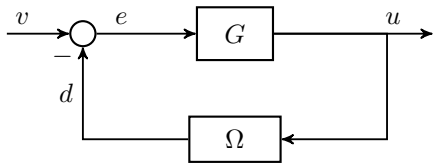
\begin{figure}[!hbt]
\centering
\begin{tikzpicture}[auto, node distance=2cm, >=stealth', line width=0.7pt]
\node[sum](sum){};
\node[block, right of= sum](P){$G$};
\node[block, below of= P, yshift=0.6cm, minimum width= 1.2cm, minimum height=0.6cm](channel){$\Omega$};
%
\draw[->]($(sum)+(-1cm,0cm)$) -- node[near start]{$v$} (sum);
\draw[->](sum) -- node[near start]{$e$} (P);
\draw[->](P) -- node[near end]{$u$} ++(2.6cm,0cm);
\draw[->](P) -- ++(2cm,0cm) |- (channel);
\draw[->](channel) -| node[near end]{$d$} node[pos=0.95]{$-$} (sum);
\end{tikzpicture}
\caption{Equivalent interconnection with input-output signals}\label{Fig:G_Omega}
\end{figure}

\section{Statistics input-output model of the channel uncertainty 
}
\label{Sec:Model_A}

In this section, the second-order statistics of the impulse response of the channel uncertainty $\Omega$ is studied. And then the relation of the statistics of its imput and output is established. 

\subsection{Second-order statistics of the channel uncertainty}

To analyze the mean-square input-output stability of the system in Fig. \ref{Fig:G_Omega},
the second-order statistics of the channel uncertainty and its output $d(k)$ are studied.
For any given instant $-\infty < n < \infty$, let the autocorrelation of the subsequence $\{\omega(k,n): -\infty< k< \infty\}$ be given by
\begin{equation}\label{Equ:Relative}
\begin{aligned}
r(l)= {\E} \Bigg\{\sum_{k = -\infty}^{\infty} \omega(k,n)\omega(k+l,n)\Bigg\}, \quad
-\infty < l < \infty.
\end{aligned}
\end{equation}
Note the fact that $\omega(k, n) \equiv 0$ for $k<n$ and $k> n+\tau$.
For the case $l=0$, only the terms with $n \leq k \leq n+\tau$ in the summation of (\ref{Equ:Relative}) may not be equal to zero.
By letting $k_1=k_2=k$, $i=k-n$ and applying Lemma \ref{Lma:omega_i}.2 into \eqref{Equ:Relative}, we obtain that
\begin{align}\label{Equ:Relative3}
r(0) 
 =\sum_{i = 0}^{\tau} \alpha_{i}^2 p_{i}(1-p_{i}).
\end{align}

According to Lemma \ref{Lma:omega_i}.3 and (\ref{Equ:Relative}), letting $k_1=k$,
$k_2=k+l$, $i_1=k-n$ and $i_2=k+l-n$ yields
\begin{align}\label{Equ:Relative2}
r(l)
=-\sum_{i = 0}^{\tau-l} \alpha_{i}\,\alpha_{i+l}\,p_{i}\,p_{i+l}, \;\; 0<l \leq \tau.
\end{align}
It holds for  $l > \tau$ that $\omega(k,n)\omega(k+l,n) \equiv 0$.
Hence, $r(l) \equiv 0$ for $l > \tau$.

In the case when $l<0$, note the fact that $ \omega(k+l,n) \equiv 0  $ for any $k+l<n$. It is verified by (\ref{Equ:Relative}) that $r(l)=r(-l)$, $\forall l<0$.

Subsequently, for any given $n$, define the energy spectral density of the subsequence
$\{\omega(k,n): -\infty< k< \infty\}$  as follows:
\begin{align}\label{Equ:Spectral_Density}
S_{\Omega}(z) = \sum_{l=-\infty}^{\infty}r(l)z^{-l}.
\end{align}

\begin{lma}\label{lemma2}
The energy spectral density $S_{\Omega}(z)$ of the channel uncertainty $\Omega$ can be written as
\begin{equation}\label{Equ:PSD_Omega}
\begin{aligned}
S_{\Omega}(z)=\frac{1}{2} \sum_{i_1,i_2=0}^{\tau}(\alpha_{i_1}z^{i_1}-\alpha_{i_2}z^{i_2})
(\alpha_{i_1}z^{-i_1}-\alpha_{i_2}z^{-i_2})p_{i_1}p_{i_2}.
\end{aligned}
\end{equation}
\end{lma}
\begin{proof}
Note the fact that any $\vert l  \vert >\tau$, $r(l)\equiv 0$. It holds that
\begin{align*}
S_{\Omega}(z) = \sum_{l=-\tau}^{\tau}r(l)z^{-l}.
\end{align*}

It follows from the definition (\ref{Equ:Spectral_Density}) of $S_{\Omega}(z)$, $r(0)$ and $r(l)$ given in
(\ref{Equ:Relative3}) and (\ref{Equ:Relative2}), respectively, that
\begin{equation}\label{Equ:X2}
\begin{aligned}
S_{\Omega}(z)=&\sum_{i=0}^{\tau} \alpha_{i}^2 p_{i}(1-p_{i})
- \sum_{\begin{subarray}{c} i_1,i_2=0\\ i_1\neq i_2  \end{subarray}  }^{\tau}  \alpha_{i_1}\alpha_{i_2} p_{i_1}p_{i_2}z^{i_1-i_2}.
\end{aligned}
\end{equation}
Note the fact that ${\displaystyle \sum_{i=0}^{\tau} p_{i}=1}$. It holds that
\begin{equation}\label{Equ:X3}
\begin{aligned}
\sum_{i=0}^{\tau} \alpha_{i}^2 p_{i}(1-p_{i})&=\sum_{i_1=0}^{\tau} \alpha_{i_1}^2
p_{i_1}\sum_{i_2=0}^{\tau} p_{i_2} - \sum_{i=0}^{\tau} \alpha_{i}^2 p_{i}^2\\
&\hspace{0cm}=\frac{1}{2}\sum_{i_1=0}^{\tau}\sum_{i_2=0}^{\tau}(\alpha_{i_1}^2+\alpha_{i_2}^2)p_{i_1} p_{i_2} - \sum_{i=0}^{\tau} \alpha_{i}^2 p_{i}^2
\end{aligned}
\end{equation}
and
\begin{equation}\label{Equ:X4}
\begin{aligned}
&\sum_{\begin{subarray}{c} i_1,i_2=0\\ i_1\neq i_2  \end{subarray}  }^{\tau}
\alpha_{i_1}\alpha_{i_2}  p_{i_1}p_{i_2}z^{i_1-i_2}
=\sum_{i_1,i_2=0}^{\tau} \alpha_{i_1}\alpha_{i_2} p_{i_1}p_{i_2}z^{i_1-i_2} - \sum_{i=0}^{\tau}\alpha_{i}^2 p_{i}^2.
\end{aligned}
\end{equation}
Substituting \eqref{Equ:X3} and \eqref{Equ:X4} into \eqref{Equ:X2} leads to \eqref{Equ:PSD_Omega}.
\end{proof}

Lemma \ref{lemma2} presents the property of  the channel uncertainty $\Omega$ in frequency-domain and the coefficients of the spectral density $S_{\Omega}(z)$. It is well-known (see for example \cite{Zhou1995}) that there exists a minimum phase polynomial $\Phi(z)$ of $z^{-1}$ with degree $\tau$ and real coefficients such
that
\begin{align}\label{S_factoriztion}
S_{\Omega}(z)=\Phi(z^{-1})\Phi(z).
\end{align}
The factorization in (\ref{S_factoriztion}) is referred to as spectral factorization of  $S_{\Omega}(z)$.

Now, the input-output relations of the channel uncertainty $\Omega$ in time-domain are studied. The following lemma is the key in analyzing these input-output relations.
\begin{lma}
\label{Lem:u_independent_tau}
Suppose that Assumptions \ref{ass1} and \ref{Assum:v} hold for the channel transmission time process
$\{\tau_k\}$ and the external input sequence $\left\{v(k)\right\}$ of the system. Then for any $k_1 \ge k_2 \ge 0$,
$\tau_{k_1}$ is independent of the channel input $u(k_1)$ and $u(k_2)$ in the system.
\end{lma}

\begin{proof}
Since the plant $P$ is assumed to be strictly proper, the controller output $u(k)$ only depends on the past inputs of $P$, which is determined by $\left\{v(0),\cdots \right.$, $\left.v(k-1)\right\}$ and $\{\tau_0,\tau_1, \cdots,\tau_{k-1} \}$, provided that $P$ and $K$ are relaxed at $k=0$.
Then by Assumptions \ref{ass1} and \ref{Assum:v}, the current channel transmission time $\tau_k$ is independent of the current and past channel inputs, which completes the proof.
\end{proof}

To study the autocorrelations of the output $d(k)$  in \eqref{omega_j_output} of the channel uncertainty
$\Omega$, denote the $i$th-component in the sum of $\tau+1$ components in the right hand side of  \eqref{omega_j_output} by $d_{i}(k)$,  $i \in {\mathcal D}$, i.e.,
$$d_{i}(k)=\omega(k, k-i)u(k-i), \quad     {i \in \mathcal{D}}.$$
Due to the fact that $\omega(k,k-i)$ is determined by $\tau_{k-i}$ only, we have the following lemma.

\begin{lma}\label{Lma:correlation_di_dj}
Suppose that Assumptions \ref{ass1} and \ref{Assum:v} hold
for the channel transmission time process $\{\tau_n\}$ and the external input sequence $\left\{v(k)\right\}$  of the system. It holds for $k_1$, $k_2=0,1,2,\cdots$ that
\begin{enumerate}
\item for $i \in \mathcal{D}$,
\begin{align*}
\E\left\{ d_{i}(k_1) d_{i}(k_2) \right\}
=\delta (k_1 - k_2) \alpha_{i}^2 p_{i}{(1 - p_{i})} \E\left\{ u^2( k_1-i ) \right\};
\end{align*}
\item   for $i_1 \ne i_2$, $i_1, i_2 \in \mathcal{D}$,
\begin{align*}
\E\left\{ d_{i_1}(k_1) d_{i_2}(k_2) \right\}
= -\delta(k_1-i_1-k_2+i_2) \alpha_{i_1} \alpha_{i_2} p_{i_1} p_{i_2} \E\left\{ u^2(k_1-i_1) \right\}.
\end{align*}
\end{enumerate}
\end{lma}
\begin{proof}
Without loss of generality, assume $k_1 > k_2$. Following Lemma \ref{Lem:u_independent_tau} and (\ref{Equ:omega_Delta}), we have that
for any $i\geq 0$, $\omega(k_1, k_1-i)$ is independent of $\omega(k_2,k_2-i)$, $u(k_1-i)$ and
$u(k_2-i)$. This leads to
\begin{align*}
&\E\left\{ d_{i}(k_1) d_{i}(k_2) \right\}\nonumber\\
=& \E\left\{\omega(k_1, k_1-i)\right\}\E\left\{ u(k_1-i) \omega(k_2, k_2-i)u(k_2-i) \right\}\\
=&0.
\end{align*}
On the other hand, it holds that
\begin{align*}
\E\left\{ d_{i}(k_1) d_{i}(k_1) \right\}
= \E\left\{\omega^2(k_1, k_1-i)\right\}\E\left\{ u^2(k_1-i) \right\}.
\end{align*}
Consequently, from Lemma \ref{Lma:omega_i}.2, Lemma \ref{Lma:correlation_di_dj}.1 holds.

Now, we prove Lemma \ref{Lma:correlation_di_dj}.2. Note that $\omega(k_1, k_1-i_1)$ is independent of
$\omega(k_2, k_2-i_2)$ for $k_1-i_1 \ne k_2-i_2$. So, it holds that
\[
\E\left\{ d_{i_1}(k_1) d_{i_2}(k_2) \right\}=0.
\]
For the case $k_1-i_1 = k_2-i_2$, $\omega(k_1,k_1-i_1)$ and $\omega(k_2, k_2-i_2)$ are independent of
$u(k_1-i_1)$, so we have that
\begin{align}\label{d_d_correlation_12}
\E\left\{ d_{i_1}(k_1) d_{i_2}(k_2) \right\}
=\E\left\{\omega(k_1, k_1-i_1)\omega(k_2, k_1-i_1)\right\}\E\left\{ u^2(k_1-i_1) \right\}.
\end{align}
Applying Lemma \ref{Lma:omega_i}.3 to (\ref{d_d_correlation_12}) leads to Lemma \ref{Lma:correlation_di_dj}.2. 
Proof is completed.
\end{proof}

The relations between the autocorrelations of the input $u(k)$ and output $d(k)$ of the channel uncertainty
$\Omega$ are presented by the following lemma.
\begin{lma}\label{d_auto}
Suppose that Assumptions \ref{ass1} and \ref{Assum:v} hold
for the channel transmission time process $\{\tau_n\}$ and the external input sequence  $\left\{v(k)\right\}$ of the system. It holds for the autocorrelations of the input $u(k)$ and output  $d(k)$ of the channel uncertainty
$\Omega$ as below:
\begin{enumerate}

\item The variances of the input $u(k)$ and output $d(k)$ satisfy that
\begin{align}\label{corrl_0}
\E\{d^2(k)\}=\sum_{i=0}^{\tau} \alpha_{i}^2 p_{i}{(1 - p_{i})}\E\{u^2(k-i)\}          ,
\end{align}

\item  The autocorrelation of $d(k)$ satisfies that for $1 \leq  |k_1-k_2| \leq \tau$,
\begin{align}\label{corrl_lp}
\E\{d(k_1)d(k_2)\}
=-\sum_{i_1,i_2=0}^{{\tau}}
\delta(k_1-i_1-k_2+i_2) \alpha_{i_1} \alpha_{i_2} p_{i_1} p_{i_2}
\E\left\{ u^2(k_1-i_1) \right\}.
\end{align}

\item It holds that for $ \vert k_1-k_2 \vert  > \tau$, $\E\{d(k_1)d(k_2)\}=0$.
\end{enumerate}
\end{lma}

\begin{proof}
The autocorrelation of $d(k)$ is determined by the autocorrelations of its components $d_i(k)$.
It follows from \eqref{omega_j_output} that
\begin{align}\label{d_d_correlation}
\hspace{-.1cm}\E\{d^2(k)\}&=\sum_{i_1=0}^{\tau}\sum_{i_2=0}^{\tau}\E\{d_{i_1}(k)d_{i_2}(k)\}\nonumber\\
\hspace{-.1cm}&=\sum_{i_1=0}^{\tau}\E\{d_{i_1}^2(k)\}+\sum_{\begin{subarray}{c} i_1,i_2=0 \\ i_1\neq
i_2\end{subarray}}^{\tau}\E\{d_{i_1}(k)d_{i_2}(k)\}
\end{align}
Applying Lemma \ref{Lma:correlation_di_dj} to (\ref{d_d_correlation}) leads to (\ref{corrl_0}), i.e.,
\begin{align*}
\E\{d^2(k)\}&=\sum_{i_1=0}^{\tau}\E\{d_{i_1}^2(k)\}
=\sum_{i=0}^{\tau} \alpha_{i}^2 p_{i}(1-p_{i}) \E\{u^2(k-i)\}.
\end{align*}

Now the autocorrelations of the sequence $\{d(k), \;k=0,1,2,\cdots \}$ are considered. For any $k_1\neq k_2$ and $\vert k_1-k_2 \vert \leq \tau$, it holds that
\begin{align}\label{d_d_correlation2}
\E\{d(k_1)d(k_2)\}
=&\sum_{i_1=0}^{\tau}\sum_{i_2=0}^{\tau}\E\{d_{i_1}(k_1)d_{i_2}(k_2)\}\nonumber\\
=&\sum_{i=0}^{\tau}\E\{d_{i}(k_1)d_{i}(k_2)\}+\sum_{\begin{subarray}{c} i_1,i_2=0 \\ i_1\neq i_2
\end{subarray}}^{\tau}\E\{d_{i_1}(k_1)d_{i_2}(k_2)\}.
\end{align}
According to Lemma \ref{Lma:correlation_di_dj}, we write (\ref{d_d_correlation2}) as (\ref{corrl_lp}).

Moreover, for any $ \vert  k_1-k_2 \vert >{\tau}$ and $i_1, i_2\in {\mathcal D}$, $\omega(k_2, k_2-i_2)$ is independent of $\omega(k_1, k_1-i_1)$. It leads to
$\E\{d(k_1)d(k_2)\}=0.$
\end{proof}

\begin{lma}
\label{corss_correl_v_d}
Suppose that Assumptions \ref{ass1} and \ref{Assum:v} hold for the channel transmission time process
$\{\tau_n\}$ and the external input sequence  $\left\{v(k)\right\}$  of the system. For any given $k_1, k_2 \ge 0$, it holds that
\begin{align}\label{v_d}
\E \{ v(k_1) d(k_2) \} =0.
\end{align}
\end{lma}
\begin{proof}
It follows from (\ref{omega_j_output}) that
\begin{align*}
v(k_1) d(k_2) =\sum_{i=0}^{\tau} v(k_1)
\omega(k_2, k_2-i) u(k_2-i).
\end{align*}
According to Lemma \ref{Lem:u_independent_tau}, Assumptions \ref{ass1} and \ref{Assum:v},
$\omega(k_2, k_2-i) $ is independent of $v(k_1)$ and $u(k_2-i)$.
Hence, (\ref{v_d}) holds.
\end{proof}

\subsection{Statistics input-output model of the channel uncertainty}

To derive the mean-square input-output stability criterion of the systems in Fig. \ref{Fig:System_G_with_uncertainty} and Fig. \ref{Fig:G_Omega}, a statistics input-output model of the channel uncertainty $\Omega$ is studied. Denote the impulse response of the nominal system $G$ by
 $g(k)$. It is assumed that the system is at rest at the initial time $k=0$. The signal $u(k)$
is given by
\begin{align*}
u(k)=\sum_{n=0}^k g(n)[v(k-n)-d(k-n)].
\end{align*}
It follows from Assumptions \ref{ass1} and \ref{Assum:v} that
$$
\E\{u(k)\}=0.
$$
Applying Lemma \ref{corss_correl_v_d}, we write the variance of $u(k)$ as below:
\begin{align}\label{var_uk1}
\E\{ u^2(k) \}= \E \left[\sum_{n=0}^k g(n)v(k-n)\right]^2
+\E  \left[\sum_{n=0}^k g(n)d(k-n) \right]^2.
\end{align}
It follows from Assumption \ref{Assum:v} that the first term in the right hand side of (\ref{var_uk1}) is given by
\begin{align}\label{var_uk2}
\E \left[\sum_{n=0}^k g(n)v(k-n)\right]^2=\sum_{n=0}^k g^2(n)\sigma_v^2(k-n).
\end{align}
On the other hand, the second term in the right hand side of (\ref{var_uk1}) is written as
\begin{align}\label{g_d_auto}
 &\E  \left[\sum_{n=0}^k g(n)d(k-n) \right]^2\nonumber\\
=&\sum_{n_1=0}^k \sum_{n_2=0}^k  g(n_1)g(n_2)\E [d(k-n_1)d(k-n_2)]\nonumber\\
=&\sum_{n_1=0}^k g^2(n_1)\E [d^2(k-n_1)]+\sum_{\begin{subarray}{c}n_1, n_2=0\\
n_1\neq n_2
\end{subarray}}^k   g(n_1)g(n_2)\E [d(k-n_1)d(k-n_2)].
\end{align}
According to Lemma \ref{d_auto}, we have
\begin{align}\label{d_1_d_2_v}
\E [d^2(k-n_1)]
=\sum_{i=0}^{\tau} \alpha_{i}^2 p_{i}{(1 - p_{i})}\E\{u^2(k-n_1-i)\}
\end{align}
and for $n_1\neq n_2$ and $\vert n_1-n_2 \vert \leq \tau$,
\begin{align*}
 &\E [d(k-n_1)d(k-n_2)]\nonumber\\
=&-\sum_{i_1,i_2=0}^{{\tau}}
\delta(-n_1-i_1+n_2+i_2) \alpha_{i_1} \alpha_{i_2} p_{i_1} p_{i_2}
\E\left\{ u^2(k-n_1-i_1) \right\}.
\end{align*}
Note the fact that $\delta(-n_1-i_1+n_2+i_2)=1$ when $n_2=n_1+i_1-i_2$, otherwise it is equal to zero. 
Hence, under the constraint $n_1 \neq n_2$,
it holds that $\delta(-n_1-i_1+n_2+i_2)=1$ when $n_2=n_1+i_1-i_2$ and $i_1 \neq i_2$, otherwise it is equal to zero.
Subsequently, it holds for $n_1\neq n_2$ and $\vert n_1-n_2 \vert \leq \tau$ that
\begin{align}\label{d_1_d_2_c}
 &\sum_{\begin{subarray}{c}n_1, n_2=0\\
n_1\neq n_2
\end{subarray}}^k   g(n_1)g(n_2)\E [d(k-n_1)d(k-n_2)]\nonumber\\
&\hspace{1cm}=-\sum_{\begin{subarray}{c}n_1, n_2=0\\
i_1\neq i_2
\end{subarray}}^k g(n_1)g(n_1+i_1-i_2)  \alpha_{i_1} \alpha_{i_2} p_{i_1} p_{i_2}
\E\left\{ u^2(k-n_1-i_1) \right\}.
\end{align}
Substituting  (\ref{d_1_d_2_v}) and (\ref{d_1_d_2_c}) into (\ref{g_d_auto})  yields that
\begin{align}\label{g_d_auto1}
 &\E  \left[\sum_{n=0}^k g(n)d(k-n) \right]^2\nonumber\\
=&\sum_{n_1=0}^k\sum_{i=0}^{\tau} g^2(n_1) \alpha_{i}^2 p_{i}{(1 - p_{i})}
\E\{u^2(k-n_1-i)\}   \nonumber\\
&-\sum_{n_1=0}^k\sum_{\begin{subarray}{c} i_1, i_2=0\\ i_1 \neq i_2 \end{subarray}}^{\tau} g(n_1)g(n_1+i_1-i_2)
\alpha_{i_1} \alpha_{i_2} p_{i_1} p_{i_2}
 \E\left\{ u^2(k-n_1-i_1) \right\}.
\end{align}

Let $l=k-n_1-i$. By change-of-variables, the first term on the right-hand side of the equality (\ref{g_d_auto1}) is written as
\begin{align}\label{Equ:Expectation_dm_dn_Term1}
&\sum_{n_1=0}^k\sum_{i=0}^{\tau} g^2(n_1) \alpha_{i}^2 p_{i}{(1 - p_{i})}
\E\{u^2(k-n_1-i)\}  \nonumber \\
=& \sum\limits_{i = 0}^{\tau}\sum\limits_{l=-i}^{k-i} g^2(k-i-l) \alpha_{i}^2 p_{i}(1-p_{i})\E\left\{ u^2(l) \right\}
\end{align}
Note the fact that $\sum\limits_{i_2=0}^{\tau} p_{i_2}=1$. The equation  (\ref{Equ:Expectation_dm_dn_Term1})  is rewritten as
\begin{align}\label{Equ:Expectation_dm_dn_Term2}
&\sum_{n_1=0}^k\sum_{i=0}^{\tau} g^2(n_1) \alpha_{i}^2 p_{i}{(1 - p_{i})}
\E\{u^2(k-n_1-i)\}  \nonumber \\
=&
\sum\limits_{i_1=0}^{\tau}\sum\limits_{l=-i_1}^{k-i_1}  g^2(k-i_1-l) \alpha_{i_1}^2 p_{i_1}\sum\limits_{i_2=0}^{\tau}p_{i_2}
\E\left\{ u^2(l) \right\}- \Xi_{k} \nonumber \\
=&
\sum\limits_{i_1, i_2=0}^{\tau}\sum\limits_{l=-i_1}^{k-i_1}  g^2(k-i_1-l) \alpha_{i_1}^2 p_{i_1}p_{i_2}
\E\left\{ u^2(l) \right\}- \Xi_{k}
\end{align}
where
$$\Xi_{k} = \sum_{i=0}^{\tau} \sum_{l=-i}^{k-i}   g^2(k-i-l) \alpha_{i}^2 p_{i}^2 \E\{u^2(l)\}. $$

On the other hand, the second term on the right-hand side of the equality (\ref{g_d_auto1}) is considered. Let
$l=k-n_1-i_1$ and it holds that
\begin{align}\label{Equ:Expectation_dm_dn_Term3}
&\sum_{n_1=0}^k\sum_{\begin{subarray}{c}i_1,i_2=0\\
i_1 \neq i_2  \end{subarray}}^{{\tau}} g(n_1)g(n_1+i_1-i_2)
\alpha_{i_1} \alpha_{i_2} p_{i_1} p_{i_2}
\E\left\{ u^2(k-n_1-i_1) \right\}\nonumber\\
=&\sum_{\begin{subarray}{c}i_1,i_2=0\\ i_1 \neq i_2 \end{subarray}}^{{\tau}}
\sum_{l=-i_1}^{k-i_1} g(k-i_1-l)g(k-i_2-l)
\alpha_{i_1} \alpha_{i_2} p_{i_1} p_{i_2}
\E\left\{ u^2(l) \right\}\nonumber\\
=& \sum\limits_{i_1,i_2=0}^{{\tau}} \sum\limits_{l=-i_1}^{k-i_1}
g(k-i_1-l)g(k-i_2-l)\alpha_{i_1} \alpha_{i_2} p_{i_1}p_{i_2}
\E\left\{u^2(l)\right\} - \Xi_{k}.
\end{align}
Substituting (\ref{Equ:Expectation_dm_dn_Term2}) and (\ref{Equ:Expectation_dm_dn_Term3}) into (\ref{g_d_auto1}) leads to
\begin{align}\label{g_d_auto2}
 &\E  \left[\sum_{n=0}^k g(n)d(k-n) \right]^2\nonumber\\
=&\sum\limits_{i_1, i_2=0}^{\tau}\sum\limits_{l=-i_1}^{k-i_1}  g^2(k-i_1-l) \alpha_{i_1}^2 p_{i_1}p_{i_2}\E\left\{ u^2(l) \right\}\nonumber\\
&-\sum\limits_{i_1,i_2=0}^{{\tau}} \sum\limits_{l=-i_1}^{k-i_1}
g(k-i_1-l)g(k-i_2-l)\alpha_{i_1} \alpha_{i_2} p_{i_1}p_{i_2}
\E\left\{u^2(l)\right\}.
\end{align}
Note the fact that $u(l)\equiv 0$, $\forall l <0$ and  $g(k-i_1-l)\equiv 0$, $\forall k-i_1-l <0$. Equality (\ref{g_d_auto2}) is rewritten as
\begin{align}\label{g_d_auto3}
 &\E  \left[\sum_{n=0}^k g(n)d(k-n) \right]^2\nonumber\\
=&\sum\limits_{i_1, i_2=0}^{\tau}\sum\limits_{l=0}^{k}  g^2(k-i_1-l) \alpha_{i_1}^2 p_{i_1}p_{i_2}
\E\left\{ u^2(l) \right\}\nonumber\\
&-\sum\limits_{i_1,i_2=0}^{{\tau}} \sum\limits_{l=0}^{k}
g(k-i_1-l)g(k-i_2-l)\alpha_{i_1} \alpha_{i_2} p_{i_1}p_{i_2}
\E\left\{u^2(l)\right\}.
\end{align}
It is verfied that
\begin{align}\label{g_d_auto4}
&\sum\limits_{i_1, i_2=0}^{\tau} [ g^2(k-i_1-l) \alpha_{i_1}^2
-g(k-i_1-l)g(k-i_2-l)\alpha_{i_1} \alpha_{i_2} ]p_{i_1}p_{i_2}\nonumber\\
=&\frac{1}{2}\sum\limits_{i_1, i_2=0}^{\tau} \left[g(k-i_1-l) \alpha_{i_1}
-g(k-i_2-l) \alpha_{i_2} \right]^2 p_{i_1}p_{i_2}
\end{align}
Hence, it is obtained from (\ref{g_d_auto3}) and (\ref{g_d_auto4}) that
\begin{align}\label{g_d_auto5}
 &\E  \left[\sum_{n=0}^k g(n)d(k-n) \right]^2\nonumber\\
=&\sum\limits_{l=0}^{k} \frac{1}{2}\sum\limits_{i_1, i_2=0}^{\tau} \left[g(k-i_1-l) \alpha_{i_1}
-g(k-i_2-l) \alpha_{i_2} \right]^2  p_{i_1}p_{i_2}
\E\left\{u^2(l)\right\}.
\end{align}
Substituting (\ref{var_uk2}) and (\ref{g_d_auto5}) into (\ref{var_uk1}) leads to
\begin{align}\label{var_uk5}
\E\{ u^2(k) \}=\sum_{n=0}^k \hat{{\mathcal  G}}(n)\sigma_v^2(k-n)
+\sum\limits_{n=0}^{k} \hat{{\mathcal  T}}(k-n)\E\left\{u^2(n)\right\}.
\end{align}
where
\begin{align}\label{T_cal_n}
\hat{{\mathcal  T}}(k-n)=\frac{1}{2}\sum\limits_{i_1, i_2=0}^{\tau}
\left[g(k-i_1-n) \alpha_{i_1}-g(k-i_2-n) \alpha_{i_2} \right]^2  p_{i_1}p_{i_2}
\end{align}
and
\begin{align}\label{G_cal_n}
\hat{{\mathcal  G}}(n)=g^2(n).
\end{align}
For convenience, denote the variance of the signal $u(k)$ by $\sigma_{u}^2(k)$.
Since $u(k)$, $k\geq 0$ is of zero-mean,
the equation (\ref{var_uk5}) is written as below:
\begin{align}\label{uk_variance}
 \sigma_{u}^2(k)
=\sum_{n=0}^k\hat{\mathcal G}(n)  \sigma_{v}^2(k-n)
+\sum_{n=0}^k\hat{\mathcal T}(n)  \sigma_{u}^2(k-n).
\end{align}
\begin{remark}
The equation (\ref{uk_variance}) is a generalization of the error variance studied in  \cite{Willems1971Frequency}. In that work, this statistics model plays a key role in developping the mean-square small gain theorem for a linear system with muplicative noises. In the next section, the mean-square input-output stability of the system is studied based  on the equation (\ref{uk_variance}).
\end{remark}

\section{Mean-square input-output stability and mean-square stability}\label{Sec:MSS}

In this section, we focus on the mean-square input-output stability criteria of the networked system over a channel with random delays. The frequency variable $z$ will be omitted whenever no confusion is caused.

It is from Parseval identity and the definition of $\mathbb{H}_2$ norm that
\begin{align}\label{T_comp}
&\lim_{k\rightarrow \infty}\sum\limits_{n=0}^{k}  \left[g(k-i_1-n) \alpha_{i_1}
-g(k-i_2-n) \alpha_{i_2}\right]^2p_{i_1}p_{i_2}  \nonumber\\
&\hspace{4cm}= p_{i_1}p_{i_2}
\left\| \left( \alpha_{i_1} z^{-i_1} - \alpha_{i_2} z^{-i_2} \right) G(z) \right\|_2^2
\end{align}
and
\begin{align}\label{G2}
\lim_{k\rightarrow \infty}\sum_{n=0}^k\hat{\mathcal{G}}(n)=\|G(z)\|_2^2.
\end{align}
Moreover, from (\ref{Equ:PSD_Omega}), (\ref{T_cal_n})  and  (\ref{T_comp}), we obtain that
\begin{align}\label{T_H2}
\lim_{k\rightarrow \infty}\sum_{n=0}^k \hat{{\mathcal  T}}(n)
=&\frac{1}{2}\sum_{i_1,i_2=0}^{\tau} p_{i_1}p_{i_2}
\left\| \left( \alpha_{i_1} z^{-i_1} - \alpha_{i_2} z^{-i_2} \right) G(z) \right\|_2^2\nonumber\\
=&\frac{1}{2\pi}\int_{-\pi}^{\pi} G^*(e^{-j\theta}) S_{\Omega}(e^{j\theta})
   G(e^{-j\theta})   d\theta.
\end{align}
Substituting (\ref{S_factoriztion}) into (\ref{T_H2}) leads to
\begin{align}\label{T_H2A}
\lim_{k\rightarrow \infty}\sum_{n=0}^k \hat{{\mathcal  T}}(n)=\|\Phi(z)G(z)\|_2^2.
\end{align}

To study the mean-square stability of the networked system, let $\vec{\sigma}_{u0:k}$ and $\vec{\sigma}_{v0:k}$ be the vectors stacked by $\sigma_{u}^2(0),\cdots, \sigma_{u}^2(k)$ and $\sigma_{v}^2(0),\cdots$,  $\sigma_{v}^2(k)$, respectively, i.e.
\begin{align}
\vec{\sigma}_{u0:k}=[\sigma_{u}^2(k),\cdots,\sigma_{u}^2(0)]^T\;\; {\rm and}\;\;
\vec{\sigma}_{v0:k}=[\sigma_{v}^2(k),\cdots,\sigma_{v}^2(0)]^T.
\end{align}
Since $\hat{\mathcal G}(0)=0$ and $\hat{\mathcal T}(0)=0$.  Let
\begin{align}\label{G_cal_k}
\hat{\mathcal G}_k
=&\begin{bmatrix} 0 &\hat{\mathcal G}(1) & \cdots  & \hat{\mathcal G}(k)\\
0 & 0  & \cdots  & \hat{\mathcal G}(k-1)\\
\vdots & \vdots & \ddots  & \vdots\\
0 & 0  & \cdots  & \hat{\mathcal G}(1)\\
0 & 0 & \cdots  &  0\\
\end{bmatrix}
,
\quad
\hat{\mathcal T}_k=
\begin{bmatrix} 0 &\hat{\mathcal T}(1) & \cdots  & \hat{\mathcal T}(k)\\
0 & 0 & \cdots  & \hat{\mathcal T}(k-1)\\
\vdots & \vdots & \ddots  & \vdots\\
0 & 0 & \cdots  & \hat{\mathcal T}(1)\\
0 & 0 & \cdots  & 0\\
\end{bmatrix}
.
\end{align}

From (\ref{uk_variance}), it holds for the vectors  $\vec{\sigma}_{u0:k}$ and $\vec{\sigma}_{v0:k}$ that
\begin{align}\label{uk_variance1}
\vec{\sigma}_{u0:k}=\hat{\mathcal G}_k  \vec{\sigma}_{v0:k} +  \hat{\mathcal T}_k \vec{\sigma}_{u0:k}.
\end{align}
Rewriting (\ref{uk_variance1}), we have that
\begin{align}\label{uk_variance2}
\vec{\sigma}_{u0:k}=(I- \hat{\mathcal T}_k)^{-1}\hat{\mathcal G}_k  \vec{\sigma}_{v0:k}.
\end{align}
Let $\hat{\mathcal S}_k=(I- \hat{\mathcal T}_k)^{-1}$. Note the facts that 
the matrix $\hat{\mathcal T}_k$ is a nonnegative matrix (i.e., its all entries are nonnegative) and upper triangular matrix. We have the lemma as follows:
\begin{lma}\label{lemma_S_k}
It holds for the matrix $\hat{\mathcal S}_k$, $k=1,2,\cdots$ that
\begin{align}\label{S_k}
\hat{\mathcal S}_{k}&=(I-\hat{\mathcal T}_{k})^{-1}
=\begin{bmatrix}
1  &  \hat{\mathcal S}(1) & \hat{\mathcal S}(2) & \cdots  &  \hat{\mathcal S}(k)  \\
0  &  1 &  \hat{\mathcal S}(1) &\cdots &  \hat{\mathcal S}(k-1)  \\
\vdots & \vdots & \vdots & \ddots  &  \vdots \\
0  &  0 & 0 &  \cdots  & 1
\end{bmatrix}
\end{align}
and 
\begin{align}\label{S_k_entry}
 \hat{\mathcal S}(k) = \hat{\mathcal T}(k) +  \hat{\mathcal S}(1) \hat{\mathcal T}(k-1)+\cdots
+ \hat{\mathcal S}(k-1)  \hat{\mathcal T}(1) 
\end{align}
with $\hat{\mathcal S}(0) = 1$  and  $\hat{\mathcal S}(1) = \hat{\mathcal T}(1)$.
\end{lma}
\begin{proof}
It follows from (\ref{G_cal_k}) that $ \hat{\mathcal T}_0=0$. Thus, we obtain that
\begin{align*}
\hat{\mathcal S}_0=(I- \hat{\mathcal T}_0)^{-1}=1,\quad \quad
\hat{\mathcal S}_1=(I- \hat{\mathcal T}_1)^{-1}
=\begin{bmatrix}
1  &   \hat{\mathcal T}(1)\\
0  &    1
\end{bmatrix}
.
\end{align*}
Let $  \hat{\mathcal S}(1) = \hat{\mathcal T}(1)$. We have
$
\hat{\mathcal S}_1
=\begin{bmatrix}
1  &   \hat{\mathcal S}(1)\\
0  &    1
\end{bmatrix}
.
$
In general, suppose that $\hat{\mathcal S}_{k}$ is given by (\ref{S_k}).
It follows from (\ref{G_cal_k})  that
\begin{align*}
\hat{\mathcal S}_{k+1}=(I-\hat{\mathcal T}_{k+1})^{-1}
&=\begin{bmatrix}
(I-\hat{\mathcal T}_{k})^{-1}  &    (I-\hat{\mathcal T}_{k})^{-1}\begin{bmatrix} \hat{\mathcal T}(k+1) \\ \vdots  \\  \hat{\mathcal T}(1) \end{bmatrix} \\
0   & 1
\end{bmatrix}
\nonumber\\
&=\begin{bmatrix}
1  &  \begin{bmatrix} \hat{\mathcal T}(1) & \cdots  & \hat{\mathcal T}(k+1) \end{bmatrix}   (I-\hat{\mathcal T}_{k})^{-1}\\
0   &  (I-\hat{\mathcal T}_{k})^{-1}
\end{bmatrix}
.
\end{align*}
Denote the $(1, k+2)$-th entry of $\hat{\mathcal S}_{k+1}$ by $\hat{\mathcal S}(k+1)$. 
Then, $\hat{\mathcal S}(k+1)$ is given by
$$
  \hat{\mathcal S}(k+1) = \hat{\mathcal T}(k+1) +  \hat{\mathcal S}(1) \hat{\mathcal T}(k)+\cdots
+ \hat{\mathcal S}(k)  \hat{\mathcal T}(1). 
$$
The matrix $\hat{\mathcal S}_{k+1}$ is written as
\begin{align*}
\hat{\mathcal S}_{k+1}&=(I-\hat{\mathcal T}_{k+1})^{-1}
=\begin{bmatrix}
1  &  \hat{\mathcal S}(1) & \hat{\mathcal S}(2) & \cdots  &  \hat{\mathcal S}(k+1)  \\
0  &  1 &  \hat{\mathcal S}(1) &\cdots &  \hat{\mathcal S}(k)  \\
\vdots & \vdots & \vdots & \ddots  &  \vdots \\
0  &  0 & 0 &  \cdots  & 1
\end{bmatrix}
.
\end{align*}
By the induction, (\ref{S_k}) and (\ref{S_k_entry}) hold for all $k \geq 0$.
\end{proof}

For any $n\times n$ matrix $A=[a_{ij}]$, denote its maximum row sum matrix norm by $\mnorm{A}{\infty}$ (see Section 5.6, \cite{Horn1986}), i.e.,
\begin{align*}
\mnorm{A}{\infty} =\max_{1\leq i \leq n}\sum_{j=1}^n |a_{ij}|.
\end{align*}
Since the matrix $\hat{{\cal S}}_k$ in  (\ref{S_k}) is a nonnegative matrix, it holds for all $k > 0$ that
\begin{align}\label{Sk_rowsum_norm}
\mnorm{\hat{\cal S}_k}{\infty} =1+\sum_{i=1}^k\hat{{\cal S}}(i).
\end{align}

\begin{lma}\label{small_gain_SS_l1}
The system in Fig. \ref{Fig:G_Omega} is mean-square input-output stable if and only if it holds that
\begin{align}\label{bound_S_infty}
\lim_{k\rightarrow \infty}\mnorm{\hat{\cal S}_k}{\infty} < \infty.
\end{align}
\end{lma}
\begin{proof}
Note (\ref{G_cal_n}) and (\ref{G2}). 
Since the nominal system $G(z)$ in Fig. \ref{Fig:G_Omega} is stable and the relative degree is greater than zero, the sums ${\displaystyle \sum_{n=1}^k\hat{{\cal G}}(n)}$, $\forall k>0$ are bounded by 
$\|G(z)\|_2^2$.  So, for any $k>0$, all entries in the vector $\hat{\mathcal G}_k  \vec{\sigma}_{v0:k}$ are bounded by the constant $\|G(z)\|_2^2 \bar{\sigma}^2$ when  the input signal $\{v(k)\}$ of the system is a  zero-mean independent process with all variances $\sigma_v^2(k)$, $k=0,1,2,\cdots$ bounded by a constant 
$\bar{\sigma}^2$.

It follows from (\ref{Sk_rowsum_norm}) that
if the inequality (\ref{bound_S_infty}) holds, there exists a constant $\bar{S}$ such that
$$
\mnorm{\hat{\cal S}_k}{\infty}<\bar{S},\quad \forall k>0.
$$ 
Hence, for the system in Fig. \ref{Fig:G_Omega}, it holds that the variance $\sigma_u^2(k)$ of the output $u(k)$ generated by the input $v(k)$ is bounded, i.e.,
$$
\sigma_u^2(k)\leq \bar{S}\|G(z)\|_2^2 \bar{\sigma}^2.
$$
That is, if  the inequality (\ref{bound_S_infty}) holds then the system Fig. \ref{Fig:G_Omega} is mean-square input-output stable. 

On the other hand, we consider the case that  the inequality (\ref{bound_S_infty}) does not hold. Without loss generlity, it is assumed that $\hat{{\cal G}}(1)\neq 0$. For any i.i.d input process $v(k)$ with variance $\sigma_v^2$ and zero-mean, it holds for all $k>0$ that all entries of the vector $\hat{\mathcal G}_k  \vec{\sigma}_{v0:k}$ except the last entry are greater than or equal to  $\hat{{\cal G}}(1)\sigma_v^2$. Following from (\ref{uk_variance2}) and Lemma \ref{lemma_S_k}, we have that
$$
\sigma_u^2(k)\geq \mnorm{\hat{\cal S}_k}{\infty} \hat{{\cal G}}(1)\sigma_v^2.
$$
Thus, it holds that 
$$
\lim_{k\rightarrow \infty}\sigma_u^2(k)\geq \lim_{k\rightarrow \infty}\mnorm{\hat{\cal S}_k}{\infty} \hat{{\cal G}}(1)\sigma_v^2 \rightarrow \infty.
$$
That is, the system is not mean-square input-output stable if the inequality (\ref{bound_S_infty}) does not hold. The proof is completed.
\end{proof}

Lemma \ref{small_gain_SS_l1} presents a criteria for the mean-square input-output stability of the system
where $\lim\limits_{k\rightarrow \infty}\mnorm{\hat{\cal S}_k}{\infty}$ is a key factor in the criteria. Note the fact that ${\hat{\cal S}_k}$ is determined by ${\hat{\cal T}_k}$. We may see from (\ref{T_H2A}) that $\lim\limits_{k\rightarrow \infty}\mnorm{\hat{\cal S}_k}{\infty}$ would be related to $\norm{\Phi(z)G(z)}{2}$ closely. To obtain  a deep understanding for the mean-square input-output stability of the networked  system, the connection between these two factors is studied. For this purpose, the following lemma is needed.

\begin{lma}{\rm (Theorem 3.50 \cite{Rudin1976})}\label{Lemma_Rudin}
Consider the sequences $\{a_k, k=0,1,2,\cdots\}$, $\{b_k,k=0,1,2,\cdots\}$ and $\{c_k=\sum\limits_{i=0}^ka_ib_{k-i}, k=0,1,2,\cdots\}$. Suppose that the sequences  $\{a_k, k=0,1,2,\cdots\}$ and $\{b_k,k=0,1,2,\cdots\}$ satisfy the conditions as follows:
\begin{enumerate}
\item $\sum\limits_{k=0}^{\infty}a_k$ converges absolutely,

\item $\sum\limits_{k=0}^{\infty}a_k=A$,

\item $\sum\limits_{k=0}^{\infty}b_k=B$.
\end{enumerate}
Then it holds that
\begin{align*}
\sum\limits_{k=0}^{\infty}c_k=AB.
\end{align*}
\end{lma}

\begin{lma}\label{S_k_sum}
Suppose the networked system in Fig. \ref{Fig:G_Omega} is mean-square input-output stable, i.e., the inequlity (\ref{bound_S_infty}) holds. Then it holds that
\begin{align}\label{Equ:S}
\sum_{k=0}^{\infty}\hat{{\cal S}}(k)  = \frac{1}{1-\norm{\Phi(z)G(z)}{2}^2}
\end{align}
and
$$
\norm{\Phi(z)G(z)}{2}<1.
$$
\end{lma}
\begin{proof}
By \eqref{S_k_entry}, it holds that
\begin{align}\label{Equ:S_K}
\hat{\cal S}(k+1) = \sum_{n=0}^{k} \hat{\cal S}(n)\hat{\cal T}(k+1-n),~~k=0,1,2,\cdots.
\end{align}
Since both $\displaystyle \sum_{n=0}^{k} \hat{\mathcal S}(n)$ and $\displaystyle \sum_{n=1}^{k} \hat{\mathcal T}(n)$ are (absolutely) convergent, it follows from Lemma \ref{Lemma_Rudin} that
\begin{align*}
\sum_{k=0}^{\infty}\sum_{n=0}^{k} \hat{\cal S}(n)\hat{\cal T}(k+1-n) & = \left[ \sum_{n=0}^{\infty} \hat{\cal S}(n)\right] \times \left[ \sum_{n=1}^{\infty} \hat{\cal T}(n)\right] \\
& = \left[ \sum_{n=0}^{\infty} \hat{\cal S}(n)\right] \times \norm{\Phi(z)G(z)}{2}^2.
\end{align*}
Noting $\hat{\mathcal{S}}(0)=1$ and adding up all entries in the both sides of the equalities in \eqref{Equ:S_K} for $k=0,1,2,\cdots$ yields 
$$
\left[ \sum_{n=0}^{\infty} \hat{\cal S}(n)\right] -1 =  \left[ \sum_{n=0}^{\infty} \hat{\cal S}(n)\right] \times \norm{\Phi(z)G(z)}{2}^2. 
$$
So (\ref{Equ:S}) holds. Moreover, note the fact that $\sum\limits_{n=0}^{\infty} \hat{\cal S}(n)>0$. This leads to 
$\norm{\Phi(z)G(z)}{2}<1$.
\end{proof}

\begin{theorem}\label{small_gain_SS}
The system in Fig. \ref{Fig:G_Omega} is mean-square input-output stable if and only if it holds that
\begin{align}\label{samll_gain_time}
 \|\Phi(z)G(z)\|_2^2   <1.
\end{align}
\end{theorem}

\begin{proof}
Suppose that the system is mean-square input-output stable. It follows from Lemma \ref{small_gain_SS_l1} 
that $\sum\limits_{k=0}^{\infty}\hat{{\cal S}}(k) <\infty$. Taking account of Lemma  \ref{S_k_sum}, we can see that  the inequality (\ref{samll_gain_time}) holds.

Now, suppose that  the inequality (\ref{samll_gain_time}) holds.
Following from (\ref{S_k}), it holds for all matrices $\hat{\cal T}_k$, $k=1,2\cdots$ that
$$
{\hat{\cal S}_k}=\sum_{i=0}^{\infty}{\hat{\cal T}_k}^i.
$$
Applying the triangle inequality and submultiplicative axioms of the matrix norm (see Section 5.6, \cite{Horn1986}) leads to
$$
\mnorm{\hat{\cal S}_k}{\infty}\leq \sum_{i=0}^{\infty}\mnorm{\hat{\cal T}_k}{\infty}^i.
$$
Taking account of (\ref{T_H2}), (\ref{G_cal_k}) and the definition of $\mnorm{\hat{\cal T}_k}{\infty}$, we have that
$$
\mnorm{\hat{\cal T}_k}{\infty}\leq  \lim_{k\rightarrow \infty}\mnorm{\hat{\cal T}_k}{\infty}
=\|\Phi(z)G(z)\|_2^2,\quad \forall k>0.
$$
So,  the inequality (\ref{samll_gain_time}) yields that
$$
\mnorm{\hat{\cal S}_k}{\infty}\leq \frac{1}{1-\|\Phi(z)G(z)\|_2^2}.
$$
This implies that $\mnorm{\hat{\cal S}_k}{\infty}$ is bounded for all $k=1,2,\cdots$. That is, the system is mean-square input-output stable.
\end{proof}

\begin{theorem}\label{corollary1}
Suppose that the system in Fig. \ref{Fig:G_Omega} is mean-square input-output stable. The input signal $v(k)$
of the system is an i.i.d process with zero-mean and variance $\sigma_{v}^2$. Then the sequence of the output variances $\sigma_u^2(k), k=0,1,2,\cdots $ is convergent and it holds that
\begin{align}\label{lim_u_sigma1}
\lim_{k \rightarrow \infty}\sigma_u^2(k)=\frac{\|G(z)\|_2^2}{1-\|\Phi(z)G(z)\|_2^2}\sigma_v^2.
\end{align}
\end{theorem}
\begin{proof}
Noting the fact that $v(k)$ is of a constant variance $\sigma_v^2$, we have that
\begin{align*}
\hat{\mathcal G}_k  \vec{\sigma}_{v0:k}&=\begin{bmatrix}\sum\limits_{n=1}^k\hat{\mathcal G}(n)\sigma_v^2,\; \cdots, \;\hat{\mathcal G}(1)\sigma_v^2 , \;0\end{bmatrix}^T
\end{align*}
and
\begin{align*}
\hat{\mathcal G}_{k+1}  \vec{\sigma}_{v0:(k+1)}&=\begin{bmatrix}\sum\limits_{n=1}^{k+1}\hat{\mathcal G}(n)\sigma_v^2,\; \cdots, \;\hat{\mathcal G}(1)\sigma_v^2, \; 0 \end{bmatrix}^T\nonumber\\
&=\begin{bmatrix}[ \hat{\mathcal G}_{k}  \vec{\sigma}_{v1:(k+1)}]^T  \;0 \end{bmatrix}^T
+\begin{bmatrix}\hat{\mathcal G}(k+1)\sigma_v^2,\; \cdots, \;\hat{\mathcal G}(1)\sigma_v^2, \; 0 \end{bmatrix}^T.
\end{align*}
Subsequently, from (\ref{uk_variance2}) and (\ref{S_k}), we obtain that
\begin{align*}
\sigma_u^2(k+1)=&\begin{bmatrix}
1  &  \hat{\mathcal S}(1) & \hat{\mathcal S}(2) & \cdots  &  \hat{\mathcal S}(k+1)
\end{bmatrix}
\hat{\mathcal G}_{k+1}  \vec{\sigma}_{v0:(k+1)}\nonumber\\
=&\begin{bmatrix}
1  &  \hat{\mathcal S}(1) & \hat{\mathcal S}(2) & \cdots  &  \hat{\mathcal S}(k+1)
\end{bmatrix}
\begin{bmatrix}[ \hat{\mathcal G}_{k}  \vec{\sigma}_{v1:(k+1)}]^T  \;0 \end{bmatrix}^T\nonumber\\
&+\begin{bmatrix}
1  &  \hat{\mathcal S}(1) & \hat{\mathcal S}(2) & \cdots  &  \hat{\mathcal S}(k+1)
\end{bmatrix}
\begin{bmatrix}\hat{\mathcal G}(k+1)\sigma_v^2,\; \cdots, \;\hat{\mathcal G}(1)\sigma_v^2, \; 0 \end{bmatrix}^T.
\end{align*}
Since $\sigma_v^2(n)=\sigma_v^2$, $\forall n$, it holds that $ \hat{\mathcal G}_{k}  \vec{\sigma}_{v1:(k+1)}= \hat{\mathcal G}_{k}  \vec{\sigma}_{v0:k} $. This leads to
\begin{align*}
\sigma_u^2(k+1)
=\sigma_u^2(k)+[\hat{\mathcal G}(k+1)+\hat{\mathcal S}(1) \hat{\mathcal G}(k) +\cdots
+\hat{\mathcal S}(k) \hat{\mathcal G}(1)]\sigma_v^2.
\end{align*}
Thus, the sequence $\{ \sigma_u^2(k) \}$ is monotonically increasing.  When the system is mena-square input-output stable, the sequence is bounded. Hence, the sequence is convergent.
Furthermore, following (\ref{uk_variance}), we have
\begin{align}\label{uk_variance_limit}
 \lim_{k \rightarrow \infty}\sigma_{u}^2(k)
= \lim_{k \rightarrow \infty}\sum_{n=0}^k\hat{\mathcal G}(n)  \sigma_{v}^2(k-n)
+ \lim_{k \rightarrow \infty}\sum_{n=0}^k\hat{\mathcal T}(n)  \sigma_{u}^2(k-n).
\end{align}
Taking account of  (\ref{G2}) and (\ref{T_H2A}), we obtain (\ref{lim_u_sigma1}).
\end{proof}

Now, the convergence of the variance sequence $\left\{\sigma_u^2(k) \right\}$ is studied for the case  when
the variance sequence $ \left\{\sigma_v^2(k) \right\}$ is convergent.

\begin{lma}\label{output_var_G}
Suppose that the nominal  system $G$ is stable and the input $v(k)$ is an independent process with zero-mean.
If the variance sequence $\left\{\sigma_{v}^2(k)\right\}$ of the input signal $v(k)$ is  convergent to $\sigma_v^2$, then the sequence ${\displaystyle \sum_{n=1}^k \hat{\mathcal G}(n)  \sigma_v^2(k-n)}$ is convergent and 
it holds that
\begin{align}\label{lim_output_Gvar}
\lim_{k \rightarrow \infty}\sum_{n=1}^{k}  \hat{\mathcal G}(n)\sigma_v^2(k-n) =\|G(z)\|_2^2 \sigma_v^2.
\end{align}
\end{lma}
\begin{proof}
Since the sequence  $\left\{\sum\limits_{n=1}^k  \hat{\mathcal G}(n)\right\} $  is convergent, the {sequence}
$\left\{\sum\limits_{n=k}^{\infty}  \hat{\mathcal G}(n)\right\} $  is convergent to zero as $k\rightarrow \infty$. Taking account of the assumption that $\left\{\sigma_v^2(k)  \right\}$ is convergent to a constant $\sigma_v^2$, we can see that for any given $\varepsilon_0 >0$,
these exists $k_0$ such that for all $k \geq k_0$, it holds that
\begin{align}\label{seq_sigma}
\left|  \sigma_v^2(k) - \sigma_v^2  \right|
<\frac{\varepsilon_0}{1+\sum\limits_{n=1}^{\infty}    \hat{\mathcal G}(n)}.
\end{align}

Let $\bar{\sigma}_{0}^2 = \max\limits_{0\le k\le k_0}|\sigma_v^2(k)-\sigma_v^2|$.
Without loss of generality, it is assumed that $\bar{\sigma}_{0}^2 > 0$.
Otherwise, $\sigma_v^2(k)$ is a constant and the convergence is apparent.
Since $G(z)$ is stable, $\sum\limits_{n=1}^\infty  \hat{\mathcal G}(n)$ is convergent to $\norm{G(z)}{2}^2$ and, there exists a sufficiently large $k_1$ ($k_1>k_0$) such that, for all $k\ge k_1 > k_0$,
\begin{align}\label{seq_G}
\sum_{n=k-k_0+1}^{k} \hat{\mathcal G}(n) \le \sum_{n=k-k_0+1}^{\infty} \hat{\mathcal G}(n) 
\le \sum_{n=k_1-k_0+1}^{\infty} \hat{\mathcal G}(n)<  \frac{\varepsilon_0}{ \bar{\sigma}_{0}^2 } .
\end{align}
For $k>k_1$, we have that
\begin{align}\label{sep}
&\left|  \sum_{n=1}^k  \hat{\mathcal G}(n) \sigma_v^2(k-n) -  \sum_{n=1}^k  \hat{\mathcal G}(n) \sigma_v^2  \right|   \nonumber\\
\le&\left|  \sum_{n=1}^{k-k_0}  \hat{\mathcal G}(n)\left[ \sigma_v^2(k-n) - \sigma_v^2\right] \right|
+
\left|  \sum_{n=k-k_0+1}^{k}  \hat{\mathcal G}(n) \left[ \sigma_v^2(k-n) -  \sigma_v^2\right] \right|.
\end{align}
It follows from (\ref{seq_sigma}) that
\begin{align}\label{seq_sigma1}
\left|  \sum_{n=1}^{k-k_0}  \hat{\mathcal G}(n)\left[ \sigma_v^2(k-n) - \sigma_v^2\right] \right|
\leq  \sum_{n=1}^{k-k_0}  \hat{\mathcal G}(n) \left|  \sigma_v^2(k-n) - \sigma_v^2 \right|
<\frac{\varepsilon_0  \sum\limits_{n=1}^{k-k_0} \hat{\mathcal G}(n)}{1+\sum\limits_{n=1}^{\infty}     \hat{\mathcal G}(n)}
<\varepsilon_0.
\end{align}
On the other hand, it holds from (\ref{seq_G}) that
\begin{align}\label{seq_G1}
\left|  \sum_{n=k-k_0+1}^{k}  \hat{\mathcal G}(n) \left[ \sigma_v^2(k-n) -  \sigma_v^2\right] \right|
&\leq  \sum_{n=k-k_0+1}^{k}  \hat{\mathcal G}(n) \left| \sigma_v^2(k-n) -  \sigma_v^2\right| \nonumber\\
&\le \max_{0\le n\le k_0-1}\{\left| \sigma_v^2(n) -  \sigma_v^2\right|\}  \sum_{n=k-k_0+1}^{k}  \hat{\mathcal G}(n)\nonumber\\
&< \bar{\sigma}_0^2 \times  \frac{\varepsilon_0}{\bar{\sigma}_0^2} = \varepsilon_0.
\end{align}
Substituting (\ref{seq_sigma1}) and (\ref{seq_G1}) into (\ref{sep}) leads to
\begin{align*}
&\left|  \sum_{n=1}^k  \hat{\mathcal G}(n) \sigma_v^2(k-n) -  \sum_{n=1}^k  \hat{\mathcal G}(n) \sigma_v^2  \right| <2\varepsilon_0.
\end{align*}
This means that
\begin{align*}
\lim_{k\rightarrow \infty}\left[ \sum_{n=1}^k  \hat{\mathcal G}(n) \sigma_v^2(k-n) -  \sum_{n=1}^k  \hat{\mathcal G}(n) \sigma_v^2  \right]=0.
\end{align*}

Since the system $G$ is stable, the sequence $\left\{\sum\limits_{n=1}^k  \hat{\mathcal G}(n) \sigma_v^2 \right\}$ is convergent. This implies that the sequence ${\displaystyle \left\{\sum\limits_{n=1}^k \hat{\mathcal G}(n)  \sigma_v^2(k-n)   \right\}}$ is convergent as $k\rightarrow \infty$ in the case when $ \sigma_v^2(k)$ is convergent. Thus, (\ref{lim_output_Gvar}) holds.
\end{proof}

\begin{theorem}
Suppose that the system in Fig. \ref{Fig:G_Omega} is mean-square input-output stable. The input signal $v(k)$
of the system is an independent process with zero-mean and bouned variances. If the  variance sequence
$\{\sigma_{v}^2(k), k=0,1,2,\cdots, \}$ of the input signal is convergent to $\sigma_v^2$, the output variance sequence $\{\sigma_u^2(k), k=0,1,2,\cdots\}$ is convergent to
\begin{align}\label{lim_u_ sigma1A}
\lim_{k \rightarrow \infty}\sigma_u^2(k)=\frac{\|G(z)\|_2^2}{1-\|\Phi(z)G(z)\|_2^2}\sigma_v^2.
\end{align}
\end{theorem}

\begin{proof}
According to (\ref{uk_variance2}) and (\ref{S_k}), we write $\sigma_u^2(k)$ as below:
\begin{align}\label{variance_u}
\sigma_u^2(k)=&
\sum_{n=1}^k \hat{\mathcal G}(n)  \sigma_v^2(k-n)  +  \hat{\mathcal S}(1) \sum_{n=1}^{k-1} \hat{\mathcal G}(n)  \sigma_v^2(k-1-n)  +\nonumber\\
&\hspace{5cm} \cdots  +  \hat{\mathcal S}(k-1)  \hat{\mathcal G}(1)  \sigma_v^2(0) .
\end{align}
Now, rewrite (\ref{variance_u}) as below
\begin{align}\label{u_k_sigma}
\sigma_u^2(k) =\sum_{n=0}^{k-1}\hat{\mathcal{S}}(n)c(k-n)
\end{align}
where $c(m) = \sum\limits_{l=1}^{m}\hat{\mathcal{G}}(m)\sigma_v^2(m-l)$. 
According to Lemma \ref{output_var_G}, the sequence $\{c(m), m=0,1,2,\cdots\}$ is bounded and  it holds that 
$$
\lim_{m\rightarrow \infty}c(m)=\lim_{m\rightarrow \infty}\sum_{l=0}^m \hat{\mathcal G}(l)  \sigma_v^2(m-l)=\|G(z)\|_2^2\sigma_v^2.
$$

On the other hand, it follows from Lemma \ref{small_gain_SS_l1} that 
\begin{align*}
\lim_{k\rightarrow \infty}\sum_{n=0}^{\infty} \hat{\mathcal S}(n) < \infty \quad\quad {\rm and} \quad\quad
\lim_{k\rightarrow \infty}\sum_{n=k+1}^{\infty} \hat{\mathcal S}(n)=0.
\end{align*}
Applying the argument used in Lemma \ref{output_var_G} for proving the convergence of the sequence ${\displaystyle \left\{\sum_{n=0}^k \hat{\mathcal G}(n)  \sigma_v^2(k-n)\right\}}$ to (\ref{u_k_sigma}), we obtain that 
\begin{align}\label{lim_sum_S}
\lim_{k\rightarrow \infty}\sigma_u^2(k)
=\left[\sum_{n=0}^{\infty}\hat{\mathcal{S}}(n)\right]\|G(z)\|_2^2\sigma_v^2.
\end{align}
Applying Lemma \ref{S_k_sum} to (\ref{lim_sum_S}), we obtain  (\ref{lim_u_ sigma1A}).
\end{proof}


Now, the stability of the networked system in Fig. \ref{Fig:G_Omega} in the mean-square sense is considered in terms of the state-space model of its nominal system $G$.
Suppose the state-space model of the nominal system $G$ given by
\begin{align}\label{state_space}
x_G(k+1)&=A_Gx_G(k)+B_Ge(k)\\
u(k)&=C_Gx_G(k)\nonumber
\end{align}
where $x_G(k)$ is the state variable of $G$, $e(k)$ and $u(k)$ are input and output of the system, respectively. The initial state of the system is $x_G(0)$. The input signal $v(k)$ is void, i.e., $v(k) \equiv 0$.

Suppose the system $G$ is stable (i.e.,  all eigenvalues of $A_G$ are within the unit circle) and its initial state $x_G(0)$ satisfies the following assumption.
\begin{assump}\label{initial_state}
The initial state $x_G(0)$ of $G$ is a random vector with zero-mean and covariance matrix $\Sigma_0$.
$x_G(0)$ is independent of the random transmission delay sequence $\left\{\tau_n, n=0,1,2\right.$, $\left.\cdots  \right\}$.
\end{assump}
\begin{theorem}
Suppose that the nominal system $G$ of the networked system  in Fig. \ref{Fig:G_Omega} is stable and its state-space model is given by (\ref{state_space}). The initial state $x_G(0)$ of the system $G$ satisfies Assumption \ref{initial_state} and the signal $v(k)\equiv 0$, $\forall k$.
The covariance of the state $x_G(k)$ approaches zero if and only if the networked system is mean-square input-output stable.
\end{theorem}
\begin{proof}
Under the assumption \ref{initial_state}, applying a similar argument for analyzing the variance $\sigma_u^2(k)$ given by (\ref{uk_variance}) to the state-space model (\ref{state_space}) yields that in the case when $v(k)\equiv 0$, the variance of $u(k)$ is given by
\begin{align}
\sigma_u^2(k)=C_GA_G^k\Sigma_0 (A_G^T)^k C_G^T
+ \sum_{n=0}^k\hat{\mathcal T}(n)  \sigma_{u}^2(k-n).
\end{align}
Hence, we obtain that $\lim\limits_{k\rightarrow \infty}\sigma_u^2(k)=0$ if and only if   $\|\Phi(z)G(z)\|_2^2   <1$.
From  (\ref{corrl_0}), it holds for the system shown in Fig. \ref{Fig:G_Omega} that the variance of the signal $d(k)$ converges to zero when the variance of the signal $u(k)$ converges to zero.

On the other hand, the system $G$ can be decomposed into four parts, i.e., controllable and observable, controllable and unobservable, uncontrollable and observable, uncontrollable and unobservable parts (see for example \cite{ChenT1999}, pp 163).  Denote the state vectors of these parts by $x_{co}$, $x_{\bar{c}o}$,
$x_{c\bar{o}}$, $x_{\bar{c}\bar{o}}$. By the property of the observability, the covariances of $x_{co}$, $x_{\bar{c}o}$ converge to zero if and only if the variance of the system's output $u$ approaches zero. Since the
uncontrollable and unobservable part disconnects with input $e$ and output $u$ of the nominal system $G$,  the covariance of its state vector $x_{\bar{c}\bar{o}}$ converges zero. The state equation of controllable and unobservable part is given by
\begin{align}\label{cobar}
x_{c\bar{o}}(k+1)=A_{c\bar{o}}x_{c\bar{o}}(k)+v_{c\bar{o}}(k)
\end{align}
where all the eigenvalues of the matrix $A_{c\bar{o}}$ are within the unit circle and $v_{c\bar{o}}$ is a linear combination of  $x_{co}$, $x_{\bar{c}o}$,
$x_{\bar{c}\bar{o}}$, $e$. Since the covariances of  $x_{co}$, $x_{\bar{c}o}$,
$x_{\bar{c}\bar{o}}$, $e$ converge to zero and the system (\ref{cobar}) is stable, the covariance of
$x_{c\bar{o}}(k)$ converges to zero. The proof is completed.
\end{proof}


\section{Mean-square stabilization via output feedback}
\label{Sec:MSS_output_feedback}

In this subsection, the control design is studied for the networked system in Fig. \ref{Fig:System_G_with_uncertainty}. Our goal is to find an optimal controller $K$ stabilizing the closed-loop system in mean-square sense.

It follows from Theorem \ref{small_gain_SS} that the control design is to design $K$ satisfying the inequality (\ref{samll_gain_time}) where $G$ is given by (\ref{Equ:Nominal_system_G(z)}). Let $\mathbb{K}$ be the set of all proper controllers $K$ stabilizing the closed-loop system $G$. The networked system is mean-square stabilizable via output feedback if and only if it holds that
$$
\min_{K\in \mathbb{K}} \|\Phi(z)G(z)\|_2^2<1.
$$

Let ${\displaystyle W=\frac{\Phi}{H}}$ which is rferred to as the frequency response of variation of the channel in \cite{SLL2021}. Then, it holds that
$$
\Phi(z)G(z)=W(z)T(z),\;\; {\rm and} \;\; T(z)=\frac{KHP}{1+KHP}
$$
where $T$ is  the complementary sensitivity function of the nominal system in Fig. \ref{Fig:System_G_with_uncertainty}.
The optimal design is to find $K^*$ such that $\|\Phi(z)G(z)\|_2^2$ achieves minimal,i.e.,
\begin{align}\label{optimal_H2}
K^*=\arg \min_{K\in \mathbb{K}} \|\Phi(z)G(z)\|_2^2=\arg \min_{K\in \mathbb{K}} \|W(z)T(z)\|_2^2.
\end{align}
That is, the optimal design is determined by the interaction between $W$ and $T$.

Now, the optimal mean-square stabilization via output feedback is studied for the networked system based on the input-output model of the system.
Let a right and left coprime factorization of the transfer function $PH$ be given by
\begin{equation*}
P(z)H(z) = N M^{-1} = \tilde{M}^{-1} \tilde{N},
\end{equation*}
where $N,M,\tilde{M},\tilde{N} \in \mathbb{RH}_\infty$ satisfy the double Bezout identity
\begin{equation*}
\begin{bmatrix}
  \tilde{V} & \tilde{U} \\
  -\tilde{N} & \tilde{M}
\end{bmatrix}\begin{bmatrix}
               M & -U \\
               N & V
             \end{bmatrix} = \begin{bmatrix}
                               M & -U \\
                               N & V
                             \end{bmatrix}\begin{bmatrix}
  \tilde{V} & \tilde{U} \\
  -\tilde{N} & \tilde{M}
\end{bmatrix}
= I
\end{equation*}
for some $U,V,\Ut,\Vt \in \mathbb{RH}_\infty$.
It is well-known that every stabilizing controller $K \in \mathbb{K}$ can be parameterized as \cite{Zhou1995}
\begin{align}
K &= (U+MQ)(Y-NQ)^{-1} \nonumber\\
&= (\Vt - Q \Nt)^{-1}(\Ut + Q\Mt),~~Q \in \mathbb{RH}_\infty. \label{Equ:Youla_param_controller}
\end{align}

Substituting \eqref{Equ:Youla_param_controller} into the complementary sensitivity function $T(z)$ yields
\begin{equation*}
T(z) = (U + MQ)\tilde{N},~~Q \in \mathbb{RH}_\infty.
\end{equation*}
The solvability of the mean-square stabilization problem is given by the following corollary.
\begin{corollary}\label{Thm:Stabilizable_condition}
The networked system in Fig. \ref{Fig:System_G_with_uncertainty} is mean-square stabilizable via output feedback if and only if
\begin{equation*}
\inf_{Q\in \rh{\infty}}   \| [(U + MQ)\tilde{N}]W\|_2^2 <1.
\end{equation*}
\end{corollary}

The optimal design to the mean-square stabilization problem is a model-matching problem. There are several methods (such as optimal $\mathbb{H}_2$ model-matching problem solver) to find the optimal solution:
$$
Q^*=\arg \inf_{Q\in \rh{\infty}}   \| (U + MQ)\tilde{N}W\|_2^2.
$$

\begin{figure}[!hbt]
\centering
\begin{tikzpicture}[auto, node distance=2cm, >=stealth', line width=0.7pt]
\node[sum](sum){};
\node[block, right of= sum, xshift=-0.4cm, yshift=0cm,  minimum width= 1cm, minimum height=0.6cm](controller){$P$};
\node[block, right of= controller](P){$K$};
\node[block, below of= P, xshift=-1.2cm, yshift=0.8cm, minimum width= 1cm, minimum height=0.6cm](channel){$H$};
\node[block, right of= P, xshift=0cm, yshift=0cm,  minimum width= 0.8cm, minimum height=0.6cm](uncertainty){$\Phi$};
\draw[->]($(sum)+(-1.2cm,0cm)$) -- node[near start]{$v$} (sum);
\draw[->](sum) -- node[near start]{$e$} (controller);
\draw[->](uncertainty) -- node[near end]{$z$} ++(1.2cm,0cm);
\draw[->](P) -- node[midway, above] {$u$}(uncertainty);
\draw[->](controller) --node[midway, above] {$y_P$} (P);
\draw[->](P) ++(1cm,0cm) |- (channel);
\draw[->](channel) -| node[near end]{$\bar{u}$} node[pos=0.9]{$-$} (sum);
\end{tikzpicture}
\caption{An equivalent LTI system with output feedback}\label{Fig:LTI}
\end{figure}
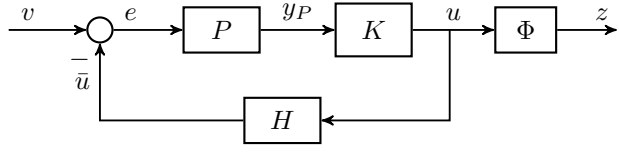

Here, the optimal design is considered in terms of the state-space model of the system.
It follows from (\ref{optimal_H2}) that the optimal design for the mean-square stabilization problem is an
$\mathbb{H}_2$ optimal control design via output feedback for the LTI system shown in Fig. \ref{Fig:LTI}.
In the system,  the whole plant ${\cal P}$ consists of three components $P$, $H$ and $\Phi$, $K$ is the output feedback controller to be designed. The goal of the control design is to find $K$ minimizing the $\mathbb{H}_2$-norm of the transfer function from $v$ to $z$, i.e., the cost function in this problem is
\begin{align}\label{H2_cost}
J=\left\| \frac{\Phi  KP}{1+KHP}  \right\|_2^2.
\end{align}
Let the state-space models of $P$, $H$ and $\Phi$ be given as below:
\begin{equation*}
P:\left\{
\begin{array}{lll}
&x_P(k+1)=A_Px_P(k)+B_Pu_P(k)\\
&\hspace{.9cm}y_P(k)=C_Px_P(k)
\end{array}
\right.
\end{equation*}
\begin{equation*}
H:\left\{
\begin{array}{lll}
&x_H(k+1)=A_Hx_H(k)+B_Hu_H(k)\\
&\hspace{.9cm}y_H(k)=C_Hx_H(k)+D_Hu_H(k)
\end{array}
\right.
\end{equation*}
\begin{equation*}
\Phi:\left\{
\begin{array}{lll}
&x_{\Phi}(k+1)=A_{\Phi}x_{\Phi}(k)+B_{\Phi}u_{\Phi}(k)\\
&\hspace{.9cm}y_{\Phi}(k)=C_{\Phi}x_{\Phi}(k)+D_{\Phi}u_{\Phi}(k)
\end{array}
\right.
\end{equation*}
where $x_P(k)$, $x_H(k)$ and $x_{\Phi}(k)$ are state variable vectors of $P$, $H$ and $\Phi$, respectively,
and $u_P(k)$, $u_H(k)$, $u_{\Phi}(k)$,  $y_P(k)$, $y_H(k)$, $y_{\Phi}(k)$ are inputs and outputs of these components, respectively.

According to the system in Fig. \ref{Fig:LTI},  it holds that $u_{\Phi}=u$, $u_P=e$, $u_H=u$, $e=-\bar{u}+v$
and $z=y_{\Phi}$. When the input signal $v(k)$ is an i.i.d process with zero-mean and unit variance, its output $z$ satisfies that
$$
\sum_{k=0}^{\infty}z^2(k)=\left\| \frac{\Phi  KP}{1+KHP}  \right\|_2^2.
$$
Denote the general plant which consists of  $P$, $H$, $\Phi$ in the system in Fig. \ref{Fig:LTI} by ${\cal P}$.
Denote the state variable of the plant ${\cal P}$ by $x(k)=[x_P^T(k)\; x_{\Phi}^T(k)\; x_H^T(k)]^T$ and the measurement of the plant by $y$. According to the structure of the system,  it holds that $y=y_P$.
The state-space model of the plant ${\cal P}$ is given by
\begin{equation*}
{\cal P}:
\left\{
\begin{array}{lll}
&x(k+1)=Ax(k)+B_1v(k)+B_2u(k)\\
&\hspace{0.7cm}z(k)=C_1x(k)+D_{12}u(k)\\
&\hspace{0.7cm}y(k)=C_2x(k)
\end{array}
\right.
\end{equation*}
where
\begin{align*}
&A=\begin{bmatrix} A_P   &   0   &  -B_PC_H  \\  0  &  A_{\Phi}  &   0 \\
0   &   0   &   A_H  \end{bmatrix},\;\;
B_1=\begin{bmatrix}B_P  \\  0   \\   0   \end{bmatrix},\;\;
B_2=\begin{bmatrix}-B_PD_H  \\  B_{\Phi}   \\   B_H   \end{bmatrix}\\
&C_1 =\begin{bmatrix} 0  &  C_{\Phi}   &   0   \end{bmatrix}, \;\;
C_2=\begin{bmatrix}C_P  &  0   &  0  \end{bmatrix}, \;\; D_{12}=D_{\Phi}.
\end{align*}
The optimal stabilization control design is the optimal $\mathbb{H}_2$ control problem to find an optimal
controller $K^*$ in minimizing the cost function $J$ in (\ref{H2_cost}). This problem is solved
by the stabilizing solutions $X$ and $Y$ to the following Riccati equations (see \cite{Chen1995Optimal}):
$$
X=A^TXA+C_1^TC_1-(C_1^TD_{12}+A^TXB_2)(D_{12}^TD_{12}+B_2PB_2)^{-1}(D_{12}^TC_1+B_2^TXA)
$$
and
$$
Y=AYA^T+B_1B_1^T-AYC_2^T(C_2YC_2^T)^{-1}C_2YA^T,
$$
respectively.

The optimal controller $K^*$ is given by
\begin{align}
\hat{x}(k+1)&=(A+B_2F+LC_2-B_2L_0C_2)\hat{x}(k)+(L-B_2L_0)y(k)  \label{Equ:Optimal_controller}\\
u(k)&=(L_0C_2-F)\hat{x}(k)+L_0y(k)\nonumber
\end{align}
where $\hat{x}$ is the state of the controller and
\begin{align*}
F&=-(D_{12}^TD_{12}+B_2^TXB_2)^{-1}(B_2XA+D_{12}^TC_1)\\
L&=-AYC_2^T(C_2YC_2^T)^{-1}\\
L_0&=FYC_2^T(C_2YC_2^T)^{-1}
\end{align*}

\section{Numerical Example}\label{section_example}
\begin{figure}[!htb]
  \centering
  \includegraphics[width=11cm]{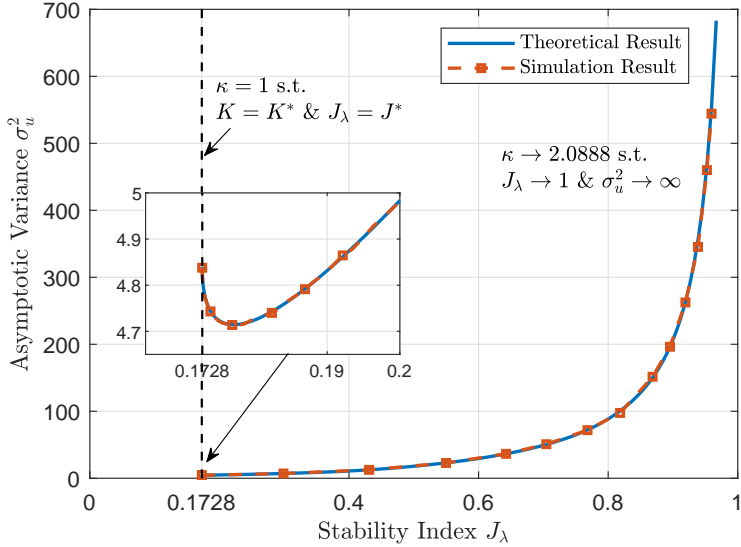}
  \caption{Stability index $J_\lambda=\left.\|\Phi G\|_2^2\right|_{K=\kappa K^*}$ v.s. the asymptotic variance of the control signal $u$}\label{Fig:WT_norm_vs_u_pow}
\end{figure}

Consider an unstable discrete-time LTI system $P$ in Fig. \ref{Fig:System_with_uncertainty}:
\begin{equation*}
\begin{aligned}
x({k+1}) &= \begin{bmatrix}
      1.2 & 0 \\
      1 & 1.1
    \end{bmatrix} x({k}) + \begin{bmatrix}
                         1 \\
                         0
                       \end{bmatrix}u({k}),\\
y({k}) &= \begin{bmatrix}
                         1 & 1
                       \end{bmatrix} x({k}).
\end{aligned}
\end{equation*}
The delay property of the unreliable channel $\Delta$ in Fig. \ref{Fig:System_with_uncertainty} is given by
\begin{align*}
\Pr\{\tau_k=0\} = 0.6,~\Pr\{\tau_k=1\} = 0.3,~\Pr\{\tau_k=2\} = 0.1.
\end{align*}
Let the signal weights of \eqref{input-output_j} be $\alpha_0 = 0.6,~\alpha_1 = 0.4$, and $\alpha_2=0$.
Then the mean channel $H$ and the spectral factorization $\Phi$ of the channel uncertainty are given by
\begin{align}\nonumber
H(z)=0.36 + 0.12z^{-1},~~\Phi(z) = 0.3188 - 0.1355 z^{-1}.
\end{align}
Note that an apparent controller in Fig. \ref{Fig:System_with_uncertainty} which probably stabilizes the closed-loop system in the mean-square sense should be the optimal controller $K^*$ designed by \eqref{Equ:Optimal_controller} that
\begin{align*}
K^*=  \frac{0.8316 z(z-1.02)}{(z-0.2)(z-0.1668)},
\end{align*}
according to which the minimum cost $J$ of \eqref{H2_cost} is
\begin{align*}
  J^*= \left.\|\Phi G\|_2^2\right|_{K = K^*} = 0.1728 <1.
\end{align*}
It follows from Theorem \ref{small_gain_SS} that the closed-loop system via the optimal controller $K^*$ is mean-square stable.
To verify this, let the controller now be $K=\kappa K^*$ with $\kappa \ge 1$ such that the cost function $J_\kappa := \left.\|\Phi G\|_2^2\right|_{K=\kappa K^*}$ would vary according to $\kappa$.
Note that $J_{\kappa=1}=J^*$.
It is expected that $J_\lambda$, which here is named as stability index, could indicate the stability of the system.
Notice that the asymptotic variance $\sigma_u^2$ of the controller output $u$ given by \eqref{lim_u_sigma1} can quantify the stability of the system, for simplicity, let the input $v(k)$ of the system be an i.i.d. process with zero-mean and unit-variance $\sigma_v^2=1$ such that
\begin{align}\label{lim_u_sigma}
\sigma_u^2 =\left.\frac{\|G(z)\|_2^2}{1-\|\Phi(z)G(z)\|_2^2}\right|_{K=\kappa K^*}.
\end{align}
It can be verified that the asymptotic variance $\sigma_u^2$ via the optimal controller $K^*$ is 4.8400, i.e., $\left.\sigma_u^2\right|_{\kappa=1}=4.8400$.
Fig. \ref{Fig:WT_norm_vs_u_pow} illustrates the numerical relationship between the stability index $J_\lambda$ and the corresponding asymptotic variance $\sigma_u^2$, by varying the implicit variable $\kappa$ from 1 to $2.0888$.
It shows that when $\kappa$ tends to $2.0888$, the stability index $J_\lambda$ would tend to 1 such that the asymptotic variance $\sigma_u^2$ would tend to infinity, which implies that the closed-loop system becomes unstable in the mean-square sense.
Since the nonlinearity between $\sigma_u^2$ and $\norm{\Phi(z)G(z)}{2}^2$, the minimum stability index does not imply the minimum asymptotic variance of the control signal, see the detail drawing in Fig. \ref{Fig:WT_norm_vs_u_pow}.

\section{Conclusion}
\label{Sec:Conclusion}

In this paper, the mean-square stability and stabilizability problem for networked systems with random data transmission delays are addressed. The delay model is studied in the time-domain and the frequency-domain.
Accordingly, the mean-square stability condition of the closed-loop system is obtained, which is a generalization of the well-known mean-square small gain theorem.
Sequentially, the connection between the mean-square input-output stability and mean-square stability of the system is studied for the networked system in terms of its input-output model and state-space model, respectively.
It is found that the mean-square input-output stability and mean-square stability is equivalent for the networked system. Moreover, an optimal design is presented for the mean-square stabilization via output feedback in terms of the state-space model of the networked system.

\end{document}